\documentclass[useAMS,usenatbib]{mn2e}

\usepackage{graphicx}
\RequirePackage{lineno}
\usepackage{amsmath}
\usepackage{lscape}
\usepackage{url}

\def\farcm{\mbox{$.\mkern-4mu^\prime$}}

\def\arcsec{\mbox{$^{\prime\prime}$}}                

\def \osix {J0610$-$2100}
\def \ten {J1024$-$0719}
\def \six {J1600$-$3053}
\def \seven {J1713$+$0747}
\def \sevenf {J1741$+$1351}
\def \twenty {J2051$-$0827}

\def \fermi {\emph{Fermi}}

\def \nsample {30}
\def \detected {six}

\def \blc {$B_\textsl{LC}$}

\def \aaa {A-type}
\def \aa {A}
\def \bbb {N-type}
\def \bb {N}
\def \ccc {W-type}
\def \cc {W}

\def \NA {five}
\def \NN {21}
\def \NW {four}
\def \NNW {25}

\title[New Fermi MSPs and the radio/gamma-ray connection]
{Six millisecond pulsars detected by the \emph{Fermi} Large Area Telescope and the radio/gamma-ray connection of millisecond pulsars
}
\author[Espinoza et al.]{
C.~M.~Espinoza,$^{1}$\thanks{E-mail: cme@jb.man.ac.uk} 
L.~Guillemot,$^{2}$\thanks{E-mail: guillemo@mpifr-bonn.mpg.de}
\"O.~\c{C}elik,$^{3, 4, 5}$
P.~Weltevrede,$^{1}$
B.~W.~Stappers,$^{1}$
\newauthor
D.~A.~Smith,$^{6}$
M.~Kerr,$^{7}$
V.~E.~Zavlin,$^{8, 9}$
I.~Cognard,$^{10, 11}$
R.~P.~Eatough,$^{2}$
\newauthor
P.~C.~C.~Freire,$^{2}$
G.H~Janssen,$^{1}$
F.~Camilo,$^{12, 13}$
G.~Desvignes,$^{2, 10, 11}$
J.W.~Hewitt,$^{3}$
\newauthor
X.~Hou,$^{6}$
S.~Johnston,$^{14}$
M.~Keith,$^{14}$
M.~Kramer,$^{1, 2}$
A.~Lyne,$^{1}$
R.~N.~Manchester,$^{14}$
\newauthor
S.~M.~Ransom,$^{15}$
P.~S.~Ray,$^{16}$
R.~Shannon,$^{14}$
G.~Theureau$^{10, 11}$
and N.~Webb$^{17, 18}$ \\
$^{1}$Jodrell Bank Centre for Astrophysics, School of Physics and Astronomy, The University of Manchester, M13 9PL, UK  \\
$^{2}$Max-Planck-Institut f\"ur Radioastronomie, Auf dem H\"ugel 69, 53121 Bonn, Germany  \\
$^{3}$NASA Goddard Space Flight Center, Greenbelt, MD 20771, USA  \\
$^{4}$Center for Research and Exploration in Space Science and Technology (CRESST), Greenbelt, MD 20771, USA  \\
$^{5}$Department of Physics and Center for Space Sciences and Technology, University of Maryland Baltimore County,\\ Baltimore, MD 21250, USA  \\
$^{6}$Universit\'e Bordeaux 1, CNRS/IN2p3, Centre d'\'Etudes Nucl\'eaires de Bordeaux Gradignan, 33175 Gradignan, France  \\
$^{7}$W. W. Hansen Experimental Physics Laboratory, Kavli Institute for Particle Astrophysics and Cosmology,\\  Department of Physics and SLAC National Accelerator Laboratory, Stanford University, Stanford, CA 94305, USA  \\
$^{8}$Space Sciences Laboratory, NASA Marshall Space Flight Center SD59, Huntsville, AL 35805, USA  \\
$^{9}$USRA Science \& Technology Institute, 320 Sparkman Drive, AL 35805, USA  \\
$^{10}$Laboratoire de Physique et Chimie de l'Environnement, LPCE UMR 6115 CNRS, F-45071 Orl\'eans Cedex 02, France  \\
$^{11}$Station de radioastronomie de Nan\c cay, Observatoire de Paris, CNRS/INSU ; F-18330 Nan\c{c}ay, France  \\
$^{12}$Columbia Astrophysics Laboratory, Columbia University, New York, NY 10027, USA  \\
$^{13}$Arecibo Observatory, HC3 Box 53995, Arecibo, PR 00612, USA \\
$^{14}$CSIRO Astronomy and Space Science, Australia Telescope National Facility, Epping NSW 1710, Australia  \\
$^{15}$National Radio Astronomy Observatory (NRAO), Charlottesville, VA 22903, USA  \\
$^{16}$Space Science Division, Naval Research Laboratory, Washington, DC 20375-5352, USA  \\
$^{17}$CNRS, IRAP, F-31028 Toulouse cedex 4, France  \\
$^{18}$GAHEC, Universit\'e de Toulouse, UPS-OMP, IRAP, Toulouse, France  \\
}

\date{\today}
\pagerange{\pageref{firstpage}--\pageref{lastpage}} \pubyear{2012}

\begin{document}

\label{firstpage}
\maketitle

\clearpage
\begin{abstract}
We report on the discovery of gamma-ray pulsations from five millisecond pulsars (MSPs) using the \fermi\ Large Area Telescope (LAT) and timing ephemerides provided by various radio observatories. 
We also present confirmation of the gamma-ray pulsations from a sixth source, PSR \twenty.
Five of these \detected\ MSPs are in binary systems: PSRs \seven, \sevenf, \six\ and the two \emph{black widow} binary pulsars PSRs \osix\ and \twenty.
The only isolated MSP is the nearby PSR \ten, which is also known
to emit X-rays.
We present X-ray observations in the direction of PSRs \six\ and
\twenty. 
While the latter is firmly detected, we can only give upper limits for
the X-ray flux of the former.
There are no dedicated X-ray observations available for the other 3 objects.

The MSPs mentioned above, together with most of the MSPs detected by \fermi, are used to put together a sample of \nsample\ gamma-ray MSPs which is used to study the morphology and phase connection of radio and gamma-ray pulse profiles. 
We show that MSPs with pulsed gamma-ray emission which is phase aligned with the radio emission present the steepest radio spectra and the largest magnetic fields at the light cylinder among all MSPs. 
As well, we also observe a trend towards very low, or undetectable,
radio linear polarisation levels.
These properties could be attributed to caustic radio emission 
produced at a range of different altitudes in the magnetosphere.
We note that most of these characteristics are also observed in the Crab pulsar, the only other radio pulsar known to exhibit phase-aligned radio and gamma-ray emission.
\end{abstract}


\section{Introduction}
Pulsed gamma-ray emission from more than one hundred pulsars has been
detected by the Large Area Telescope (LAT) aboard the \fermi\ 
{\sl Gamma-ray Space
  Telescope}\footnote{https://confluence.slac.stanford.edu/display/GLAMCOG/
Public+List+of+LAT-Detected+Gamma-Ray+Pulsars} \citep{aaa+10c,Fermi2PC}. 
About a third of them are millisecond pulsars (MSPs), fast rotators
that spin down very slowly and steadily.
MSPs are thought to have been spun-up through accretion of material
from an evolved companion star \citep{acrs82,rs82}.
During accretion, the system appears as a low-mass X-ray binary
(LMXB) and no pulsed radio emission is observed.
Although other possibilities have been considered \citep[e.g.][]{rud91}, it is commonly believed that accretion is responsible for the low surface magnetic fields observed amongst MSPs. 
In this model, the original field has been buried by the
in-falling material, decreasing its surface strength and decreasing the
magnetic torque acting on the neutron star \citep[][]{rom90,bv91}.

Most MSPs are found in binary systems.
Among those in binaries, there is a growing number in very tight binary
systems with low mass companions, the so called \emph{Black Widows} \citep[e.g.][]{rob11}.
It is believed that most solitary MSPs came through the black widow
evolutionary path, in which the companion is ablated by the pulsar wind, either during the X-ray phase or once the pulsar turns on in the radio \citep{el88,rste89}.

The gamma-ray emission mechanism for MSPs and normal pulsars is believed to be the same \citep{aaa+09}.
MSPs exhibit the lowest gamma-ray luminosities among gamma-ray pulsars, a consequence of their low spin-down energy rates ($\dot{E}$), compared to  young pulsars \citep{aaa+10c}.
Most observed gamma-ray pulse profiles consist of two dominant, sharp peaks,
which suggest the emission is caustic in nature \citep{aaa+10c}.
Among the different models for radio and gamma-ray emission of
pulsars, these properties tend to favour models in which gamma rays
are generated in the outer magnetosphere
\citep[e.g.][]{ry95,dr03} and radio emission is produced at lower
altitudes \citep[e.g.][]{ran93}.
This is supported by the fact that most gamma-ray detected radio pulsars exhibit gamma-ray pulses arriving out of phase with the radio pulsations \citep{aaa+10c}.
However, this scheme has been challenged by the increasing number of MSPs found to exhibit phase-aligned radio and gamma-ray pulse profiles,
suggesting that, at least in these cases, they are both produced at a similar location in the magnetosphere \citep{aaa+10b,faa+11,gjv+12}.
It has been noted that some of the MSPs showing phase-aligned radio and gamma-ray emission exhibit very low levels of radio linear polarisation \citep{gjv+12}, which is predicted to some extent by caustic emission models \citep{vjh12}. 

Before these discoveries, the Crab pulsar was the only pulsar known to
exhibit phase-aligned radio and gamma-ray emission \citep{khwf03}.
MSPs present radio polarisation properties similar to normal radio pulsars \citep{xkj+98,ymv+11,kjb+12} but often much wider pulse profiles.
Interpulses and bridge emission between different peaks are more common among MSPs \citep{ymv+11}.
MSPs present radio spectra with a slope similar to normal radio pulsars, although arguably slightly steeper \citep{tbms98,kxl+98,mkkw00a}.

As well as detecting many new radio MSPs in gamma rays \citep{aaa+09},
\fermi\ has also pointed the way to many MSP discoveries via their
gamma-ray emission properties \citep[e.g.][]{rrc+11,cgj+11,kjr+11,gfc+12,kcj+12}.  
Here we report on another \detected\ MSPs detected in gamma rays by
the \fermi\ LAT, all
of which were previously known radio pulsars (PSRs \osix, \ten, \six,
\seven, \sevenf\ and \twenty).
This is the first time that pulsed gamma-rays are detected from these
sources, except for PSR \twenty, for which a 4\,$\sigma$ detection was
reported by \citet{wkh+12}.

By combining these \detected\ MSPs with 24 previously reported gamma-ray MSPs we are able to start evaluating trends in the emission properties of the gamma-ray MSP	sample.
In particular, we use this sample to study the pulse profile
properties and other connections between the radio and gamma-ray
characteristics of these objects. 

In section 2 we describe the methods used to detect and analyse the gamma-ray pulsations from the \detected\ new gamma-ray MSPs and in section \ref{detections} we comment on their multi-wavelength properties.
 In section \ref{orbsearch} we present a search for orbital modulation in the gamma-ray emission of the five MSPs in binary systems.
The study of the radio and gamma-ray properties of most known gamma-ray MSPs is presented in section \ref{morfanalys} and the discussion of the study is in section \ref{discussion}.
A summary of our work is in section \ref{summary}.

\section{Methodology: detecting gamma-ray pulsations}
\label{methods}
The detection of pulsed gamma-ray emission from known MSPs is possible
through the use of precise rotational ephemerides obtained through
frequent radio observations, which are used to
assign a rotational phase to each gamma-ray photon \citep{sgc+08}.
Photons are then binned in pulse phase to create a histogram which
represents the light curve of the pulsar.

\subsection{Radio analysis}
We use pulse times of arrival (TOAs) obtained from radio observations
with the Arecibo Observatory \citep[AO,][]{fbw+11}, the Nan{\c c}ay
Radio Telescope \citep[NRT,][]{tch+05}, the Jodrell Bank Observatory
\citep[JBO,][]{hlk+04}, the Parkes Observatory \citep[PKS,][]{wjm+10} and
the Westerbork Synthesis Radio Telescope \citep[WSRT,][]{ksv08,vkv+02}. 
See section \ref{timing} for more details on the use of the different datasets. 

\begin{table*}
\begin{minipage}{126mm}
\caption{Properties of the radio ephemerides used to fold the gamma-ray data. \label{ta:parfiles}}
\begin{tabular}{llrrccc}
\hline
\multicolumn{1}{c}{Pulsar} & \multicolumn{1}{c}{Observatories} &  \multicolumn{1}{c}{$N_\textrm{TOAs}$} & 
\multicolumn{1}{c}{RMS} & \multicolumn{1}{c}{MJD range} & 
\multicolumn{1}{c}{DM} & \multicolumn{1}{c}{$\delta\phi$} \\
& & & \multicolumn{1}{c}{$\mu$s} & & \multicolumn{1}{c}{pc\,cm$^{-3}$} & \multicolumn{1}{c}{$10^{-3}$} \\
\hline
J0610$-$2100 & JBO, NRT, PKS	&	111	& 2.84	& 54509 -- 55850 & 60.67(1)$^\dagger$   & 6    \\
J1024$-$0719 & JBO, NRT, WSRT	&	145	& 1.82	& 54590 -- 55837 & 6.488(1)    & 0.4  \\
J1600$-$3053 & NRT         	&	193	& 1.20	& 54564 -- 55797 & 52.3218(4)  & 0.2  \\
J1713+0747 & JBO, NRT, WSRT	&	257	& 0.74	& 54501 -- 55804 & 15.9929(1)  & 0.1	\\
J1741+1351 & AO         	&	8192	& 0.77	& 52840 -- 55889 & 24.2014(1)  & 0.04	\\
J2051$-$0827 &	JBO, WSRT	&	1590	& 14.52	& 54338 -- 55880 & 20.73673(2) & 0.01	\\
\hline
\end{tabular}
{\sc Note.---} Columns contain the observatories involved in the observations, the total number of TOAs used ($N_\textrm{TOAs}$), 
 the root mean square variation (RMS) of the timing residuals and the time range of validity for the timing solution.
 The last two columns show the DM values and the uncertainty of the phases assigned to the gamma-ray photons 
 caused by the uncertainties on the DM, $\delta\phi$.
DM uncertainties on the last quoted digits are in parentheses.
These uncertainties correspond to the formal errors given by {\sc tempo2} and may under-represent
the real uncertainties.
The photon phase errors ($\delta\phi$) caused by the uncertainty on the DM were calculated using
these values.
$^\dagger$Value taken from \citet{bjd+06}.
\end{minipage}
\end{table*}


To study the phase alignment between gamma and radio
pulses it is necessary to account for the delay suffered by the radio
waves in their passage through the interstellar medium.
In order to do this precisely, accurate estimates of the dispersion
measure (DM) are essential, especially for rapidly rotating pulsars
\citep{sgc+08}.
With the aim of having accurate and updated DM values, we measured the
DM using multi-frequency radio data taken during the time of the \fermi\ mission for five of the \detected\ gamma-ray MSPs (Table \ref{ta:parfiles}).

The best available radio TOAs were used to produce radio ephemerides
valid from at least 3 months before the start of the \fermi\ mission
until October 2011.
Ephemerides were produced by analysing the TOAs with the timing
software {\sc Tempo2} \citep{hem06}. 
Because of the use of different pulse-profile templates to match the 
observations, hence different points of reference, and the
existence of small clock divergences between observatories, the
different datasets sometimes require phase-alignment.  
Thus, in addition to the spin frequency and its first derivative,
binary parameters, position, proper motion and DM, we also fitted for
phase delays between the datasets from different observatories. 
More information on the timing solutions built for each
pulsar can be found in Table \ref{ta:parfiles} and section \ref{timing}.

While pulsar timing can provide most of the known parameters describing pulsars, distances can be obtained by different techniques.
If no better distance estimate was available (e.g. via parallax), the DM was used to calculate a distance based on the Galactic Free Electron Model \citep[NE2001,][]{cl02}. 
The best distance estimates ($d$) for the 6 MSPs are listed in Table~\ref{ta:mainradio}.
We note that the error bars of the DM-based distances could potentially  be largely underestimated, affecting the error bars quoted for gamma-ray luminosities and efficiencies (section \ref{gamma-sp}).

Because MSPs present low rotational period derivatives ($\dot{P}$),
the Shklovskii effect \citep{shk70} can be significant for those
objects with a large proper motion (like PSR \ten, see
below).
We have corrected all $\dot{P}$ values for this effect
and the corrected values can be found in Table~\ref{ta:mainradio}. 
Other radio properties like position, rotational and orbital
periods and proper motion are also found in Table~\ref{ta:mainradio}.

\subsubsection{Building timing solutions, dispersion measures and
  distances} 
\label{timing}
For PSR \osix\ most observations during the last 4 years have been
carried out at $\sim 1.4$\,GHz, hence it is not possible to perform a 
sensitive analysis of the DM during the time of the \fermi\ mission.
Nonetheless, using a few $\sim 2$\,GHz TOAs and the small
spread of observing frequencies around $1.4$\,GHz of the other
available TOAs, we are confident that the DM has not changed by more
than 0.02\% with respect to the value quoted in the radio discovery paper \citep{bjd+06}. 
This ensures an accuracy better than $0.006$ rotations on the phases
assigned to the gamma-ray photons.

For PSR \ten, we 
measure a DM of $6.488\pm0.001$~pc\,cm$^{-1}$, which
indicates a distance of $380\pm40$\,pc, based on the NE2001 model.
However, the LutzÐ-Kelker-bias-corrected distance is $500\pm100$\,pc \citep{hbo06,vwc+12}.
As is discussed below, considerations of the transverse movement
and the Shklovskii effect are important before adopting any of
these values. 
There are slightly different published proper motion values for this
pulsar \citep[see][]{bjb+97,hbo06}. 
Using 5 years of NRT data at $\sim1.4$\,GHz we fitted for position
and proper motion and obtained values entirely consistent with those
in \citet{hbo06}. 
The large total proper motion measured ($\sim59$~mas\,yr$^{-1}$) implies
a significant contribution to the observed $\dot{P}$ value by the
Shklovskii effect.
In fact, it seems that most of the observed spin-down comes from
this effect. 
The parallax distance indicates a negative intrinsic $\dot{P}$ and the maximum possible distance, in order to obtain a positive $\dot{P}$, is 410\,pc. 
Using the lowest possible distance given by the parallax uncertainties (i.e. 400\,pc), we calculate the upper limit $\dot{P}\leq0.05\times10^{-20}$.
We will use this $\dot{P}$ upper limit and adopt the LutzÐ-Kelker-bias-corrected distance calculated by \citet{vwc+12}.

For PSR \six,
the timing solution was obtained using data at two main frequencies
($1.4$ and $2.1$\,GHz), that allowed us to calculate an accurate and
up-to-date DM. 
We also included the first time derivative for the DM in the fit (i.e. it was allowed to vary linearly with time), measuring a value consistent with that in \citet{yhc+07}. 

For PSR \seven,
a distance of $0.9\pm0.1$\,kpc is obtained from the NE2001 model 
using our DM estimate (Table \ref{ta:mainradio}).
However, using 12 years of radio timing, \citet{sns+05} measured this pulsar's yearly parallax and calculated a distance of $1.1\pm0.1$\,kpc, 
which is the value we adopt here.

For PSR \sevenf,
the timing solution was obtained with
observations carried out at the AO using the L-wide receiver, recording
data between $1.1$ and $1.6$\,GHz, which offers enough bandwidth to
monitor DM variations. 
Based on the NE2001 model and the DM obtained from these observations
the distance to this pulsar would be $0.9\pm0.1$\,kpc. 
The same dataset contains the clear effects of a yearly
parallax, suggesting a similar distance of 
$\sim1.08\pm 0.05$\,kpc \citep[this is a preliminary value,][]{paulo2012},
which is the value we use.

For PSR \twenty,
the orbital parameters are changing rapidly: \citet{sbm+98} 
reported the measurement of the first time derivative of the
orbital period, \citet{dlk+01} detected up to 3 time
derivatives of the orbital period and \citet{lvt+11} detected
variations of the orbital period and the projected semi-major axis.
Almost 4 years of WSRT data at $1.4$\,GHz was used to fit for position,
proper motion and binary parameters, using the ELL1 binary model
\citep{lcw+01} implemented in {\sc Tempo2}, and including one
derivative of the orbital period. 
A second WSRT dataset taken at $0.35$\,GHz and JBO TOAs at $1.5$\,GHz
were then used together with the $1.4$\,GHz data to fit for the DM and
its first derivative, keeping all orbital and astrometric parameters
fixed. 

These timing solutions will be made available through the
\emph{Fermi} Science Support Center\footnote{
  http://fermi.gsfc.nasa.gov/ssc/data/access/lat/ephems/}. 
%
\begin{table*}
\begin{minipage}{151mm}
\caption{Main pulsar parameters for the 6 detected MSPs.}
\label{ta:mainradio}
\begin{tabular}{lcccccccccl}
\hline
\multicolumn{1}{c}{Pulsar} 
& \multicolumn{1}{c}{$\ell$} 
& \multicolumn{1}{c}{$b$} 
& \multicolumn{1}{c}{$P$} 
& \multicolumn{1}{c}{$\dot{P}$} 
& \multicolumn{1}{c}{$P_B$}
& \multicolumn{1}{c}{$\mu$} 
& \multicolumn{1}{c}{$d$}
& \multicolumn{1}{c}{$\dot{P}_{\rm corr}$} 
& \multicolumn{1}{c}{$\dot{E}_{\rm corr}$} 
& \multicolumn{1}{c}{Refs.} \\
\multicolumn{1}{c}{} & \multicolumn{1}{c}{\small(deg)} & \multicolumn{1}{c}{\small (deg)} & \multicolumn{1}{c}{\small (ms)} & \multicolumn{1}{c}{\small$10^{-20}$} 
& \multicolumn{1}{c}{\small days} & \multicolumn{1}{c}{\small mas yr$^{-1}$} & \multicolumn{1}{c}{\small kpc}
& \multicolumn{1}{c}{\small$10^{-20}$} & \multicolumn{1}{c}{\small$\times 10^{33}$\,erg\,s$^{-1}$} & \multicolumn{1}{c}{} \\
\hline
J0610$-$2100	&	227.8	&	-18.2	&	3.86	&     1.24	&	0.3	&  18.2(2)  & 4(1)$^\dagger$ &	0.1(3)  & 1(2)    & 1,	2	\\
J1024$-$0719$^*$&	251.7	&	40.5	&	5.16	&     1.85	&	--	&  59.9(2)  & 0.5(1)       &	$<0.05$	& $<0.1$  & 1,	3	\\
J1600$-$3053	&	344.1	&	16.5	&	3.60	&     0.95	&	14.4	&  7.2(3)   & 2(1)         &	0.84(4) & 7.1(4)  & 4,	3	\\
J1713$+$0747	&	28.75	&	25.2	&	4.57	&     0.85	&	67.8	&  6.30(1)  & 1.1(1)       &	0.805(4)& 3.33(2) & 5		\\
J1741$+$1351	&	37.89	&	21.6	&	3.75	&     3.02	&	16.3	&  11.71(1) & 1.08(5)      &    2.89(1) & 21.68(4)& 6       	\\
J2051$-$0827	&	39.19	&	-30.4	&	4.51	&     1.27	&	0.1	&  7.3(4)   & 1.0(2)       &	1.21(1) & 5.21(5) & 7,	2	\\
\hline
\end{tabular}
{\sc Note.---} The first 8 columns contain pulsar names, positions in galactic coordinates, periods, period derivatives, 
orbital periods, proper motions and distances.
The 9$^{th}$ and 10$^{th}$ columns give the values for the period derivative and spindown energy rate, 
both corrected for the Shklovskii effect ($P_{\rm corr}$ and $\dot{E}_{\rm corr}$).
Uncertainties in the last quoted digits are in parentheses. 
References in the last column are for proper motion and distance: 
(1)		{This work};
(2)		NE2001 model \citep{cl02};
(3)		\citet{vwc+12} 
(4)		\citet{vbc+09};
(5)		\citet{sns+05};
(6)             \citet[preliminary,][]{paulo2012}
(7)		\citet{lvt+11}.
$^\dagger$See a discussion about this value at the end of section \ref{gamma-sp}.
$^*$See a discussion about the distance to this MSP and the
  Shklovskii corrections in sections \ref{timing} and \ref{gamma-sp}.
\end{minipage}
\end{table*}

\subsection{Gamma-ray analysis}
\label{gammapart}

To study the gamma-ray emission from these six MSPs we selected \fermi\ LAT data taken between 2008 August 4 and 2011 August 4 using the \fermi\ Science Tools (STs)\footnote{http://fermi.gsfc.nasa.gov/ssc/data/analysis/scitools/overview.html}. We restricted the dataset to events with energies between 0.1 and 100 GeV, reconstructed directions within 15$^\circ$ of the pulsar locations, zenith angles smaller than 100$^\circ$, and belonging to the ``source'' class of the P7\_V6 instrument response functions (IRFs). We rejected the data collected when the LAT rocking angle exceeded 52$^\circ$, when the instrument was not operating in the science observation (or configuration) mode or when the data quality flag was not set as good.

\subsubsection{Spectral properties}
\label{gamma-sp}
The gamma-ray spectral properties of the pulsars were determined using a binned maximum likelihood method, as implemented in the \emph{pyLikelihood} Python module of the \fermi\ STs.
This method fits a model representing the point sources in the selected Region Of Interest (ROI) and the diffuse emission to the data, and finds the best-fit parameters optimising the likelihood function describing the data. Our models included all sources from the 2FGL catalog \citep{Fermi2FGL} found within 20$^\circ$ of the pulsar positions. Parameters of sources within 8$^\circ$ of the pulsars were left free in the fit, while parameters of sources more than 8$^\circ$ away were fixed at the values listed in the 2FGL catalog. The diffuse Galactic  emission was modelled using the \emph{gal\_2yearp7v6\_v0} map cube. The residual instrument background and the diffuse extragalactic emission were modelled using the \emph{iso\_p7v6source} template\footnote{These diffuse models are available for download from the \emph{Fermi} Science Support Center, see \url{ http://fermi.gsfc.nasa.gov/ssc/data/access/lat/BackgroundModels.html}}. The normalisation of the diffuse components were left free in the fits. The first step of the spectral analysis involved modelling the pulsar spectra with exponentially cutoff power-law (ECPL) shapes of the form:
\begin{eqnarray}
\frac{dN}{dE} = N_0 \left( \frac{E}{1\ \mathrm{GeV}}\right)^{-\Gamma} \exp \left( - \frac{E}{E_c} \right),
\end{eqnarray}
where $N_0$ is a normalisation factor, $\Gamma$ is the photon index and $E_c$ is the cutoff energy of the spectrum. The spectral parameters for PSRs J0610$-$2100 and J1713+0747 found with this model are listed in Table~\ref{gammaprop}, along with the derived integrated photon fluxes $F$ and energy fluxes $G$ above 0.1 GeV.   
These pulsar spectra were also fitted with a simple power-law model to test the validity of the ECPL model by comparing the goodness of the fit for the two models. 
For both pulsars the power-law model was rejected with more than 3\,$\sigma$ significance.
For the other four pulsars, fitting the spectra with all three parameters of the ECPL spectral shape left free was unsuccessful and yielded unsatisfactory results indicative of bad convergence issues. 
To estimate the cutoff energy and the integrated energy flux $G$ for these pulsars, a second spectral fit was performed using a ECPL model with two free parameters only, fixing $\Gamma$ to a value of 1.3, which is the average seen for the 32 strongest MSPs in the second catalog of LAT pulsars  \citep[][in preparation]{Fermi2PC}.
The results of these fits are also listed in Table \ref{gammaprop}. In this table, the first quoted uncertainties are statistical, and the second are systematic. For PSRs J0610$-$2100 and J1713+0747, the latter uncertainties were estimated by running the fitting procedure using bracketing IRFs where the effective area was perturbed by the estimated uncertainties $\pm$10\% at 0.1 GeV, $\pm$5\% near 0.5 GeV, and $\pm$10\% at 10 GeV, using linear extrapolations (in log space) in between \citep{aaa+12c}.
For the other four MSPs, the systematic errors caused by fixing the photon index to a nominal value of $1.3$ probably dominate the errors due to uncertainties in the IRFs. 
To estimate the systematic uncertainties resulting from this choice, we fitted the data using photon indices of $0.7$ and $2$, these values representing the extrema observed for 32 strong gamma-ray MSPs \citep{Fermi2PC}. The best-fit parameters for $\Gamma = 0.7$ and $\Gamma = 2$ provide the limits on the actual values listed in Table \ref{gammaprop}.
We note that the fits obtained with $\Gamma = 2$ were inconsistent with the data in all cases, which resulted in very large values of $E_c$ for PSRs J1600$-$3053 and J1741+1351 and, therefore, large systematic errors for these objects.
Finally, the gamma-ray luminosities $L_\gamma = 4 \pi f_\Omega G d^2$ and the efficiencies $\eta = L_\gamma / \dot E$ for the conversion of spin-down luminosity into gamma-ray radiation were calculated assuming a beaming factor of $f_\Omega = 1$ \citep[see][for more details]{Watters2009} which is common for outer magnetospheric emission models \citep[e.g.][]{Venter2009}. These quantities are given in Table~\ref{gammaprop}.

Two objects present anomalous efficiencies.
We obtain $\eta>0.8$ for PSR \ten, determined by the maximum $\dot{E}$ allowed by the $\dot{P}$ value adopted in section \ref{timing}, which is very low. 
However, if this MSP was at 350\,pc, $\dot{E}$ would be larger and thus the efficiency would be smaller and closer to values commonly observed 
(see the discussion on this MSP's distance in section \ref{timing}).
We note that this problem could also be alleviated if $f_\Omega$ was smaller than $1$.
By modelling the gamma-ray emission of a few MSPs, \citet{Venter2009} calculated $f_\Omega$ for different geometric configurations (magnetic inclination and line of sight). 
Although for known gamma-ray MSPs they find $f_\Omega$ values close to 1, under some specific geometric configurations their calculations show that $f_\Omega$ also could be very small.

The other case is the large efficiency obtained for PSR J0610$-$2100 ($\eta\sim10$, Table \ref{gammaprop}).
An efficiency above 1 is unphysical and to have $\eta < 1$ requires $d^2 f_\Omega < 1$\,kpc$^2$.
As noted above, it is possible that the flux correction factor is not exactly $f_\Omega=1$, but smaller.
We also consider the possibility that the large DM-based distance (Table \ref{ta:mainradio})
is due to material along the line of sight, not modelled in NE2001. 
Infrared images acquired by WISE (4.6, 12, 22 microns) and by IRAS  (25, 60, 100 microns) show pronounced nebulosity around PSR J0610$-$2100. 
A distance of 1\,kpc in this direction corresponds to 15\,pc\,cm$^{-3}$ in NE2001. 
Typical cloud sizes and over-densities can accommodate this DM discrepancy. 
For a distance of 1\,kpc (taking $f_\Omega = 1$), the gamma-ray luminosity would be as low as $0.8\times10^{33}$\,erg\,s$^{-1}$, the Shklovskii-corrected spin-down power would be $6\times10^{33}$\,erg\,s$^{-1}$ and the efficiency would be $\eta=0.1$\,.
A smaller distance also improves the transverse velocity estimate.
The measured proper motion (Table \ref{ta:mainradio}) and a distance of 4\,kpc imply a transverse velocity $V_\text{T}=345$\,km\,s$^{-1}$, a rather large value, given that the mean value for MSPs has been estimated to be $\sim90$\,km\,s$^{-1}$, with a dispersion of 20\,km\,s$^{-1}$  \citep{lml+98}.
With a distance of 1\,kpc the transverse velocity of PSR J0610$-$2100 would be $\sim85$\,km\,s$^{-1}$, very similar to the mean value for MSPs.
However, we note that if $f_\Omega<1$ the distance (thus the transverse velocity) could be larger.
For example, if we impose the maximum value of $V_\text{T}$ allowed from the dispersion \citep[i.e. $V_\text{T}=110$\,km\,s$^{-1}$,][]{lml+98} 
 then $d=1.3$\,kpc and $f_\Omega<0.6$.

\subsubsection{Search for pulsations}
It has been shown that pulsation searches can be made more sensitive by weighting each event by the probability that it originates from the considered gamma-ray point source \citep{Kerr2011,gjv+12}. These event probabilities depend on the spectra of the sources of interest and of the other sources in the region. The best-fit spectral models and the \fermi\ ST \texttt{gtsrcprob} were thus used to compute the event probabilities. The event arrival times were finally phase-folded with the \fermi\ plug-in distributed in the \textsc{Tempo2} pulsar timing package \citep{hem06,Ray2011}. The weighted $H$-test statistics \citep{Kerr2011} obtained by reducing the regions of interest to 5$^\circ$ around the pulsars are listed in Table~\ref{gammaprop}. These test statistics values all correspond to pulsation significances larger than 5\,$\sigma$. We therefore have detected pulsed gamma-ray emission from PSRs J0610$-$2100, J1024$-$0719, J1600$-$3053, J1713+0747 and J1741+1351 for the first time, and confirmed the marginal detection of PSR J2051$-$0827 presented by \citet{wkh+12}.

Probability-weighted gamma-ray light curves for the six MSPs are shown in Figures 1 to 6, along with radio and X-ray profiles when available. The gamma-ray background levels in these figures were obtained by summing the probabilities that the events are not due to the pulsar, as described in \citet{gjv+12}. Statistical error bars were calculated as $\sqrt{\sum_i w_i^2}$, where $w_i$ is the event probability and $i$ runs over events in a given phase bin \citep{Pletsch2012}. In Table~\ref{gammaprop} we list the radio-to-gamma-ray lags $\delta$ and gamma-ray peak separations $\Delta$ for pulsars with multiple gamma-ray peaks, where the positions of the gamma-ray peaks were determined by fitting the integrated light curves above 0.1 GeV with Lorentzian functions, and the positions of the radio peaks were defined as the maxima of the radio light curves.

\begin{table*}
\begin{minipage}{157mm}
\caption{Gamma-ray light curve and spectral properties for the 6 detected MSPs.
\label{gammaprop}}
\begin{tabular}{lcccccc}
\hline
Parameter & J0610$-$2100$^d$ & J1024$-$0719 & J1600$-$3053 \\
\hline
Weighted $H$-test statistics        & 50.872 & 45.405 & 103.794 \\ 
Pulsation significance ($\sigma$)   & 6.05   & 5.69   & 8.84    \\ 
Gamma-ray peak multiplicity         & 1      & 1      & 1       \\ 
Radio-to-gamma-ray lag, $\delta$    & 0.57 $\pm$ 0.01 & 0.6 $\pm$ 0.2 & 0.17 $\pm$ 0.02 \\ 
Gamma-ray peak separation, $\Delta$ & ---  & ---             & ---             \\ 
Photon index, $\Gamma$ & 1.2 $\pm$ 0.4 $_{-0.1}^{+0.1}$  & 1.3$\dagger$ & 1.3$\dagger$ \\
Cutoff energy, $E_c$ (GeV) & 1.6 $\pm$ 0.8 $_{-0.2}^{+0.3}$  & 2.2 $\pm$ 0.7 $_{-0.9}^{+4.1}$  & 5.0 $\pm$ 1.8 $_{-2.4}^{+42.4}$ \\
Photon flux, $F$ ($\geq$ 0.1 GeV, $10^{-9}$ ph cm$^{-2}$ s$^{-1}$) & 7.8 $\pm$ 2.5 $_{-0.2}^{+0.2}$  & 4.0 $\pm$ 1.1 $_{-1.4}^{+3.4}$  & 3.5 $\pm$ 0.9 $_{-1.3}^{+3.0}$  \\
Energy flux, $G$ ($\geq$ 0.1 GeV, $10^{-12}$ erg cm$^{-2}$ s$^{-1}$) & 6.6 $\pm$ 1.1 $_{-0.3}^{+0.4}$  & 3.8 $\pm$ 0.7 $_{-0.5}^{+0.8}$  & 5.2 $\pm$ 1.0 $_{-0.4}^{+0.7}$  \\
Luminosity, $L_\gamma / f_\Omega$ ($\geq$ 0.1 GeV, $10^{33}$ erg s$^{-1}$) & 10 $\pm$ 6 $_{-6}^{+6}$  & 0.11 $\pm$ 0.05 $_{-0.05}^{+0.05}$  & 4 $\pm$ 3 $_{-3}^{+3}$ \\
Efficiency, $\eta / f_\Omega$ ($\geq$ 0.1 GeV) & 11 $\pm$ 27 $_{-11}^{+27}$  & $>$0.8 $\pm$ 0.3 $_{-0.3}^{+0.3}$  & 0.5 $\pm$ 0.4 $_{-0.4}^{+0.4}$  \\
\hline
Parameter  & J1713+0747 & J1741+1351 & J2051$-$0827 \\
\hline
Weighted $H$-test statistics        & 53.675 & 51.662 & 51.718 \\
Pulsation significance ($\sigma$)   & 6.23   & 6.10   & 6.10   \\
Gamma-ray peak multiplicity         & 1      & 1      & 1      \\
Radio-to-gamma-ray lag, $\delta$    & 0.32 $\pm$ 0.05 & 0.74 $\pm$ 0.01 & 0.54 $\pm$ 0.04 \\
Gamma-ray peak separation, $\Delta$ & ---    & ---    & --- \\
Photon index, $\Gamma$& 1.6 $\pm$ 0.3 $_{-0.2}^{+0.1}$  & 1.3$\dagger$ & 1.3$\dagger$\\
Cutoff energy, $E_c$ (GeV) & 2.7 $\pm$ 1.2 $_{-0.3}^{+0.3}$  & 3.1 $\pm$ 1.6 $_{-1.3}^{+11.7}$  & 2.0 $\pm$ 0.5 $_{-0.8}^{+4.4}$ \\ 
Photon flux, $F$ ($\geq$ 0.1 GeV, $10^{-9}$ ph cm$^{-2}$ s$^{-1}$) & 13.3 $\pm$ 3.7 $_{-0.2}^{+0.2}$  & 2.8 $\pm$ 1.1 $_{-1.0}^{+2.5}$  & 4.3 $\pm$ 1.0 $_{-1.3}^{+2.5}$ \\ 
Energy flux, $G$ ($\geq$ 0.1 GeV, $10^{-12}$ erg cm$^{-2}$ s$^{-1}$) & 10.2 $\pm$ 1.5 $_{-0.4}^{+0.3}$  & 3.1 $\pm$ 0.8 $_{-0.4}^{+0.7}$  & 3.8 $\pm$ 0.7 $_{-0.3}^{+0.4}$ \\ 
Luminosity, $L_\gamma / f_\Omega$ ($\geq$ 0.1 GeV, $10^{33}$ erg s$^{-1}$) & 1.5 $\pm$ 0.3 $_{-0.3}^{+0.3}$  & 0.43 $\pm$ 0.12 $_{-0.07}^{+0.11}$  & 0.5 $\pm$ 0.2 $_{-0.2}^{+0.2}$ \\ 
Efficiency, $\eta / f_\Omega$ ($\geq$ 0.1 GeV) & 0.44 $\pm$ 0.10 $_{-0.08}^{+0.08}$  & 0.020 $\pm$ 0.005 $_{-0.003}^{+0.005}$  & 0.10 $\pm$ 0.03 $_{-0.03}^{+0.03}$ \\ 
\hline
\end{tabular}
{\sc Note.---}{
The first quoted uncertainties are statistical, while the second are systematic. 
Values marked with the symbol $\dagger$ were fixed in the spectral
analysis and in these cases the systematic errors were calculated in a different manner.
Details on the measurement of these parameters are given in section
\ref{gammapart}.  
For PSR~J1024$-$0719, the reported gamma-ray efficiency $\eta$ is a
lower limit, since only an upper limit on the Shklovskii-corrected
period derivative, $\dot P$, is known (sections \ref{timing} and \ref{gamma-sp}). 
$^d$Please see the discussion about this pulsar's efficiency in section \ref{gamma-sp}.
}
\end{minipage}
\end{table*}


\section{Multi-wavelength properties of the six MSPs}
\label{detections}
Below we present some basic information on these MSPs and 
detailed descriptions of their pulse profiles, in gamma rays and radio.
If X-ray observations were available, descriptions of their main X-ray properties are also given.

\subsection{PSR J0610$-$2100}
\label{0610X}
 
\begin{figure}
\begin{center}
\includegraphics[width=75mm]{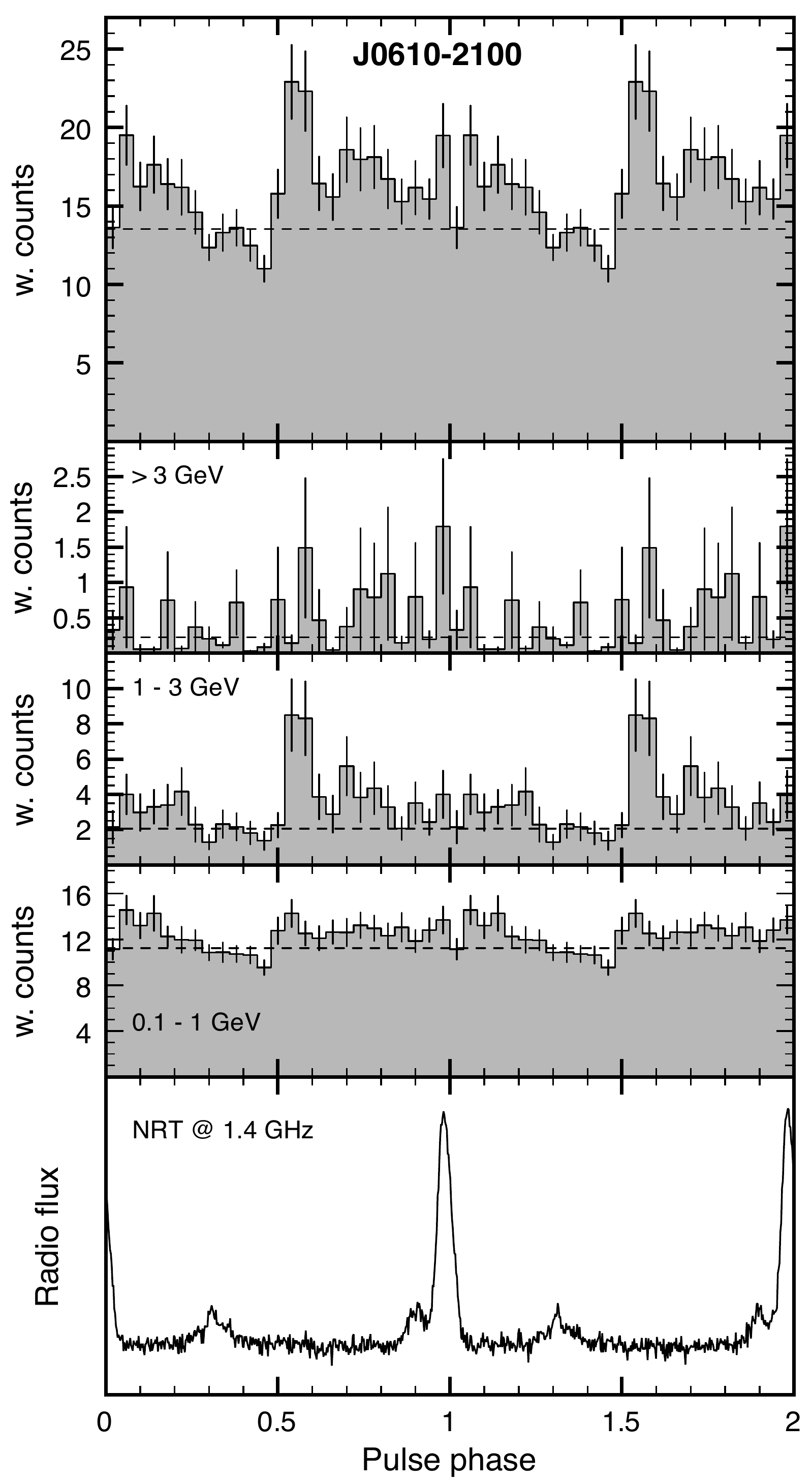}
\caption{
Gamma-ray and radio pulse profiles for PSR \osix.
The profiles are shown twice, here and elsewhere in the paper, to elucidate their complex nature and to clearly visualise alignment or otherwise.
The gamma-ray light curve for events with energies above $0.1$\,GeV is shown in the top panel. The following panels show the light curve for different gamma-ray energy bands, as indicated.  
The histograms were made using photon weights, and the horizontal dashed lines represent the estimated background emission (section \ref{methods}). 
The bottom plot shows the radio pulse profile at the indicated frequencies.} 
\label{0610lc} 
\end{center} 
\end{figure}

PSR \osix\ is a $3.8$\,ms pulsar in a $6.9$-hour orbit with a $\sim0.02$\,M$_{\odot}$ white dwarf \citep{bjd+06}.
Given the low mass of the companion and the short orbital period, PSR \osix\ is one of the \emph{Black Widow} binary systems. 
However, in this case there are no radio eclipses nor DM variations
caused by ablation of the companion star \citep{bjd+06}, that are
usually associated with these systems \citep[e.g. PSR
B1957$+$20,][]{fst88}. 
Pulsed gamma-ray emission seems to come mostly from the $1$--$3$\,GeV band (Fig. \ref{0610lc}).

In X-rays, the position of PSR \osix\ has been imaged with the {\it
  Swift} XRT instrument in six observations performed in March 2010,
April and May 2011. 
The effective exposures of those observations were very short, between $0.4$ and $4.1$\,ks, and with the pulsar position offset from the XRT aim point by angles between $0\farcm 2$ and $4\farcm 6$.  
PSR \osix\  was not detected with {\it Swift}. 
We used the level 2 data taken in Photon Counting mode from these six data sets ($10.0$\,ks of the total effective exposure)
and applied the approach developed by \citet{wwt+07} for
statistical estimates on source detections to put a 3\,$\sigma$ limit
on the XRT source count rate of $1.3$\,cnt\,ks$^{-1}$ (in a 
$30$\arcsec-radius aperture centred at the pulsar radio position). 
Assuming, for simplicity,  a power-law X-ray spectrum with a photon 
index $\Gamma_\text{X}=2$ and absorbing hydrogen column density  $N_{\rm H}=0.9\times 10^{21}$\,cm$^{-2}$
(equal to the total Galactic H$_{\rm I}$ column in the pulsar
direction\footnote{http://heasarc.gsfc.nasa.gov/cgi-bin/Tools/w3nh/w3nh.pl}), this count rate translates into an 3\,$\sigma$ upper limit of $1.1\times 10^{32}\,(d/4.0\,{\rm kpc})^2$\,erg\,s$^{-1}$ for the X-ray luminosity of the pulsar in the band $0.3$--$10$\,keV.
This limit would be more stringent if the distance is over-estimated (see the end of section \ref{gamma-sp}).

PSR \osix\ has a $1.4$-GHz pulse profile which consists of 3
components: a main pulse and precursor separated by 0.1 rotations
and another component which lags the main pulse by about 0.3
rotations (see Fig. \ref{0610lc}).

\subsection{PSR J1024$-$0719}
This is a $5.2$\,ms isolated pulsar located less than $500$\,pc from the Sun \citep{bjb+97}.
An X-ray counterpart was proposed by \citet{bt99} based on {\it ROSAT} observations and later,
using XMM-Newton data, \citet{zav06} detected pulsed emission.
A candidate optical counterpart was reported by \citet{srr+03} but
the association could be due to positional coincidence and further
observations are necessary.

The gamma-ray pulse profile of PSR \ten\ exhibits one 
broad gamma-ray peak, roughly half a rotation wide, with its centre
preceding the radio peak by about $0.5$ rotation.  
Pulsed emission seems to be present mostly in the 1--3\,GeV band.

\begin{figure}
\begin{center}
\includegraphics[width=75mm]{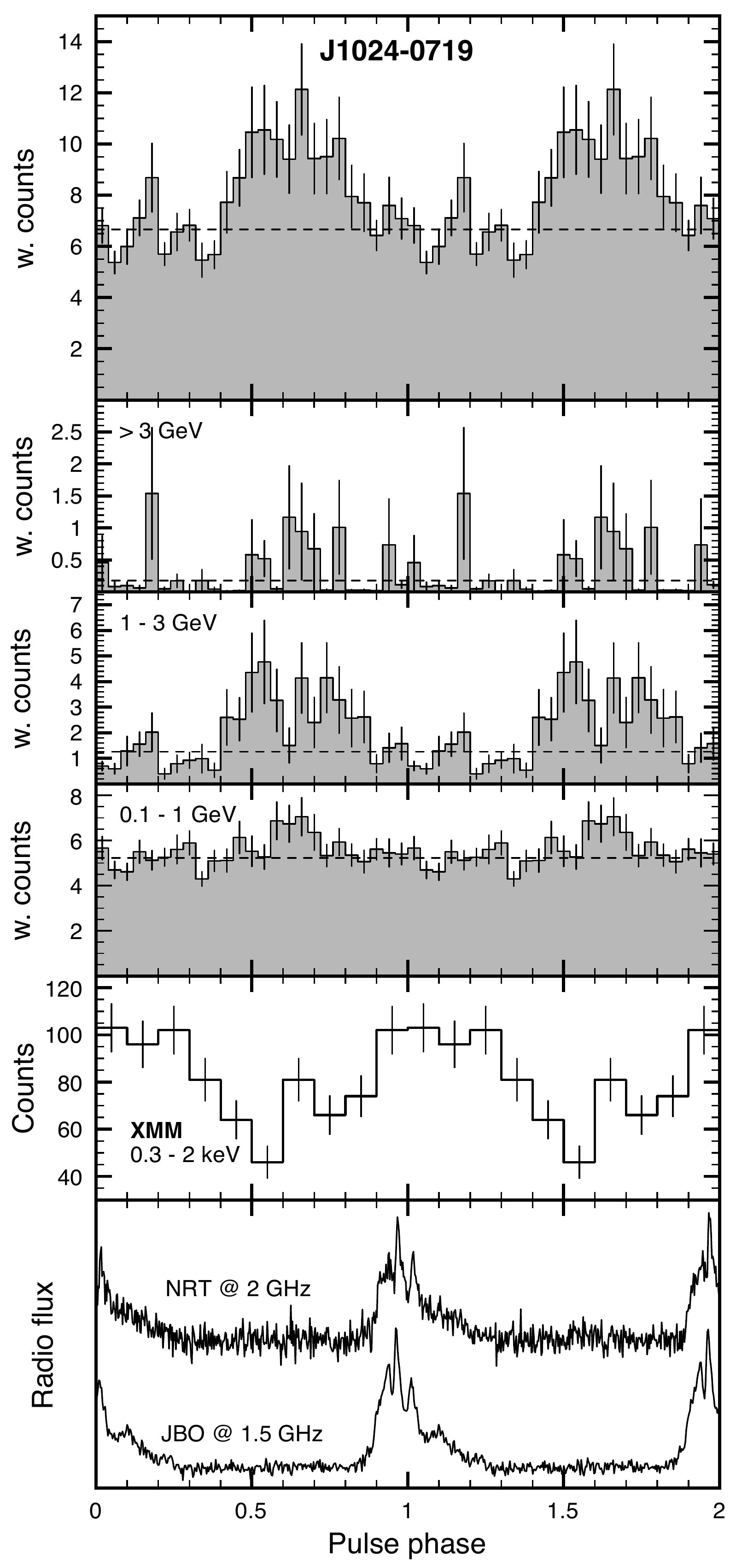}
\caption{Gamma-ray and radio pulse profile for PSR \ten\ (see Fig.~\ref{0610lc} for details). 
The X-ray light curve is presented on the fifth panel from top to bottom.}
\label{1024lc}
\end{center} 
\end{figure}

We include the X-ray pulse profile of PSR \ten\ in Fig. \ref{1024lc}
obtained from data collected with the EPIC-pn instrument, operated
in Timing mode, in a XMM-{\it Newton} observation conducted in December 2003 for a 66-ks effective exposure \citep{zav06}.  
We processed the data as in \citet{zav06} but using the latest XMM-{\it Newton} data reduction software (SAS v. 11.0.0).
The light curve was obtained from $815$ events extracted from 
columns $38$--$39$ in the one-dimensional EPIC-pn CCD image and in the
$0.3$--$2$\,keV energy range to maximise the signal-to-noise ratio. 
We used the {\it photons} plugin{\footnote{
    http://www.physics.mcgill.ca/$\sim$aarchiba/photons\_plug.html}} 
for {\sc tempo2} to assign a phase to each selected photon, together
with a timing ephemeris obtained from JBO observations between
November 2002 and October 2004.

The H-test reports a $7.5$\,$\sigma$ detection of the pulsed emission,
confirming the result of \citet{zav06}. 
The estimated intrinsic pulsed fraction, corrected for the background
contribution, is $54\pm 20$\%.  
The X-ray pulse peaks at phase $0.1$--$0.15$, as determined by fitting a series of harmonics to the pulse profile, and its shape indicates a predominantly thermal origin of the pulsar X-ray emission \citep[see][for more details]{zav06}.

At $\sim1.4$\,GHz, this pulsar presents a complex average radio pulse profile consisting of about 9 components spanning $0.2$ rotations,
which emerge from a wider base almost half of a rotation wide.
The profile at $2$\,GHz looks very similar (Fig. \ref{1024lc}).
The emission at $1.4$\,GHz is almost completely linearly polarised in the leading part of the pulse, involving the three main peaks, but it is not polarised on the trailing part \citep{ymv+11}. 
There is little position angle (PA) variation detected across the profile.

\subsection{PSR J1600$-$3053}
This is a 3.6 ms pulsar in a  14.4 day orbit with its binary companion \citep{jbo+07}.
Its gamma-ray peak is $\sim0.3$ rotations
wide and exhibits a sharp leading edge and a slowly decaying trailing
edge. 
Most pulsed emission comes from the two higher energy bands, although
some pulsed emission at earlier phases may also come from the 
$0.1$--$1$\,GeV band.

PSR \six\ was observed with XMM-{\it Newton} in February 2008 with the EPIC-MOS and EPIC-pn instruments, operated in Full Window and Timing mode (respectively), for $\sim30$-ks effective exposures.  
Data were reduced with the SAS, version 11.0.0. 
Examining the EPIC-MOS images around the radio position yielded no detection of the pulsar. 
The emission detected in the one-dimensional EPIC-pn image was heavily contaminated by enhanced background and by another bright source in the field of view.
Therefore, despite the much higher sensitivity of the EPIC-pn instrument,  no timing analysis of these data turned out to be meaningful \citep[see the example of PSR J0034$-$0534 in][]{zav06}.
Using the approach described in section \ref{0610X}, we put a 3\,$\sigma$ upper limit of $1.2$\,cnt\,ks$^{-1}$ on the pulsar EPIC-MOS count rates, in 
the $0.3$--$10$\,keV band (as measured from a $30$\arcsec-radius aperture centred at the pulsar radio position).   
For the power-law X-ray model of $\Gamma_\text{X}=2$ and absorbing hydrogen
column density  $N_{\rm H}=1.0\times 10^{21}$\,cm$^{-2}$,  these count
rates translate into a rather deep 3\,$\sigma$ upper limit of
$7.1\times 10^{30}\,(d/2\,{\rm kpc})^2$\,erg\,s$^{-1}$ for the X-ray luminosity, in the $0.3$--$10$\,keV band.

\begin{figure}
\begin{center}
\includegraphics[width=75mm]{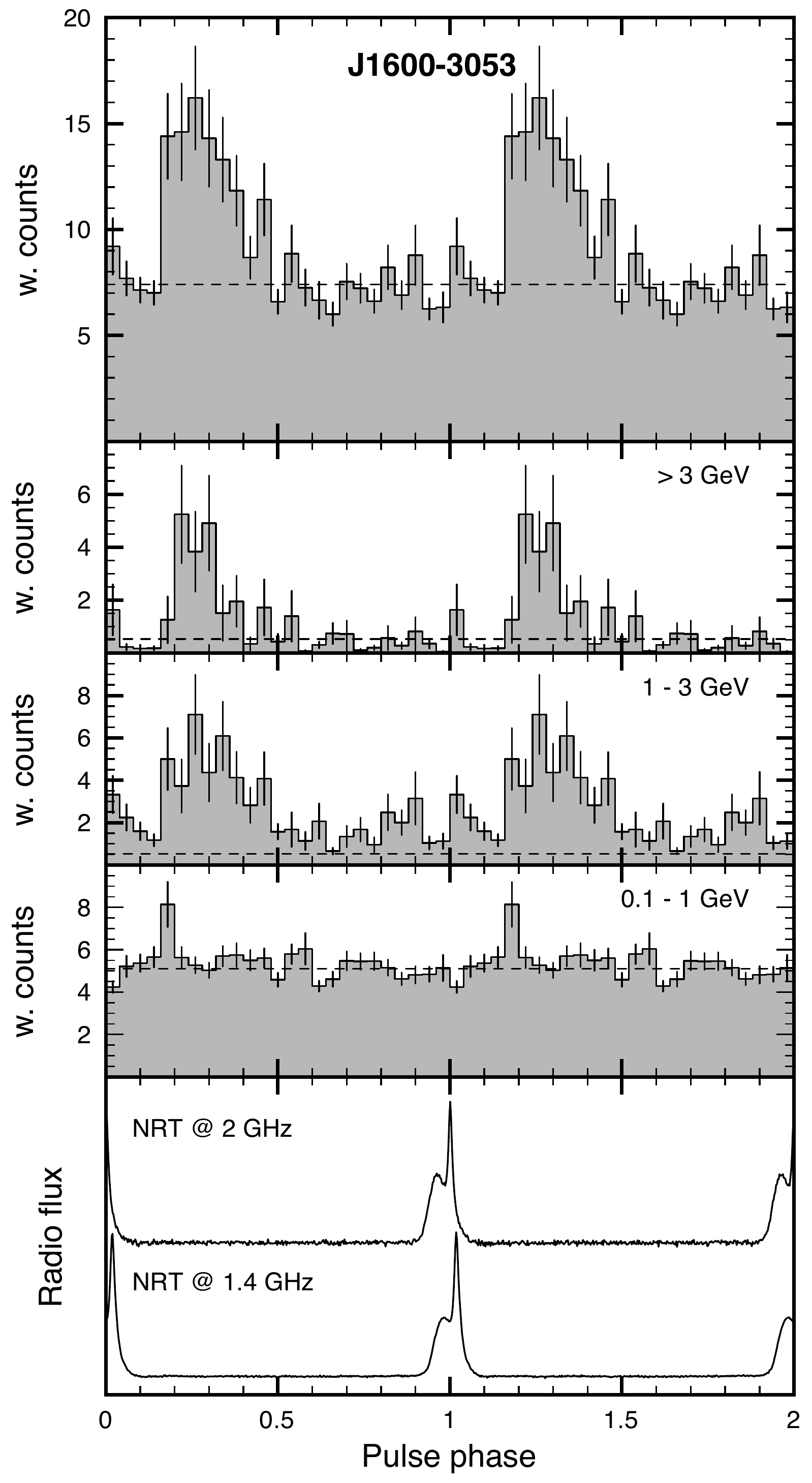}
\caption{Gamma-ray and radio pulse profile for PSR \six\ (see Fig. \ref{0610lc} for details).} 
\label{1600lc}
\end{center}
\end{figure}

The average pulse profile at $1.4$\,GHz is dominated by a sharp pulse
which is preceded by a broader additional component, less than 0.1 rotations
earlier, with roughly half its amplitude.
There are no major differences with the pulse profile at 2\,GHz (Fig. \ref{1600lc}).
The emission at $1.4$\,GHz is 30\% linearly polarised and exhibits
two orthogonal PA jumps, the second one coincident with a sense reversal of the circular polarisation \citep{ymv+11}.

\subsection{PSR J1713+0747}
\begin{figure}
\begin{center}
\includegraphics[width=75mm]{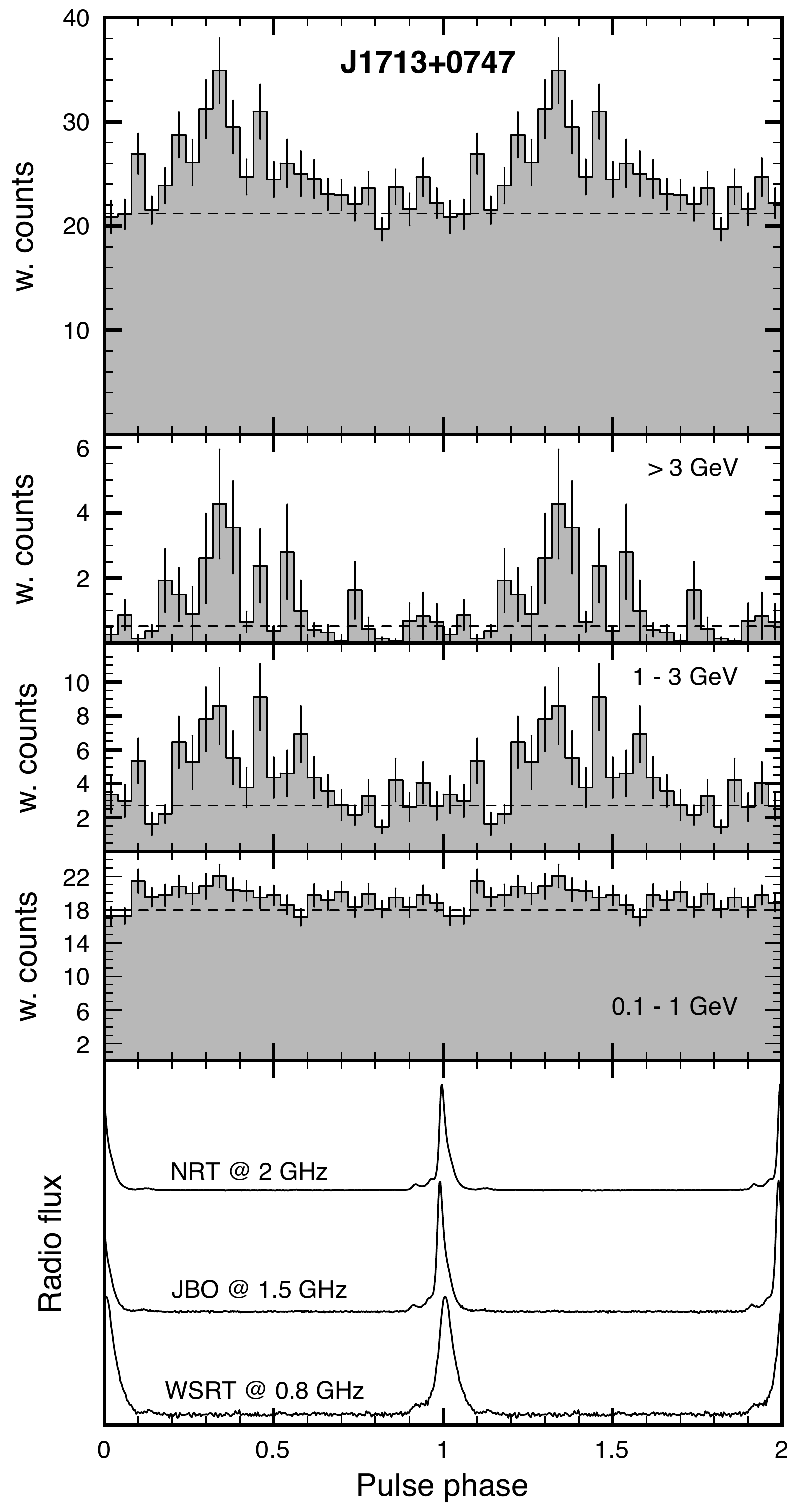}
  \caption{Gamma-ray and radio pulse profile for PSR \seven\ (see Fig. \ref{0610lc} for details).}
\label{1713lc}
\end{center}
\end{figure}

This is a $4.6$\,ms pulsar in a $68$\,day orbit with a white dwarf \citep{fwc93}.
The gamma-ray pulse profile of PSR \seven\ is wide and roughly
triangular but not exactly symmetric, with a shallower trailing edge
(Fig. \ref{1713lc}). 

In the X-ray band, the position of PSR \seven\ has only been observed in four short exposures by {\it Swift}/XRT in Photon Counting mode 
($3.1$\,ks of the total effective exposure). 
Using these XRT level 2 data and the same approach as in section \ref{0610X}  yielded a 3\,$\sigma$ limit of $1.9\times 10^{31}\,(d/1.1\,{\rm kpc})^2$\,erg\,s$^{-1}$ on the X-ray luminosity of this MSP in $0.3$â--$10$\,keV (adopting $N_{\rm H}=0.5\times10^{21}$\,cm$^{-2}$).

The average pulse profile of PSR \seven\ at $1.5$\,GHz consists of one
sharp pulse with a weak trailing component and two small components on
its base, preceding the main pulse by 0.04 and 0.08 rotations \citep{ymv+11}.
Beside some broadening at lower frequencies, there are no major differences between the pulse profile at $0.8$ and $2$\,GHz.
There is a third, very shallow component visible in the three radio bands trailing the main peak by a little more than 0.1 rotations (see Fig. \ref{1713lc}).
The emission at $1.4$\,GHz is almost 100\% linearly polarised at the leading and trailing edges of the profile.
There are two orthogonal PA jumps at each side of the main peak and a third one just before the shallow trailing component.
The second jump (and possibly also the first) is coincident with a sense reversal of the circular polarisation \citep{ymv+11}.

\subsection{PSR J1741+1351}
This is a 3.8 ms pulsar in a 16 day period binary system \citep{jbo+07}.
The gamma-ray peak of PSR \sevenf\ spans $\sim0.3$ rotation and exhibits relatively sharp edges.
Most of the emission is in the two upper energy bands and leads the radio peak, an unusual situation among gamma-ray pulsars. Other cases like this are the MSPs PSRs J1744$-$1134 and J2124$-$3358 \citep[e.g.][]{Fermi2PC}.

PSR J1741+1351 was in the field of view of the {\it Swift} XRT instrument
in January 2012 for $3.7$\,ks. However, as it was projected very close to
the edge of the XRT image (with a $7\farcm3$ offset angle), no
meaningful limit for the pulsar X-ray flux could be derived.

\begin{figure}
\begin{center}
\includegraphics[width=75mm]{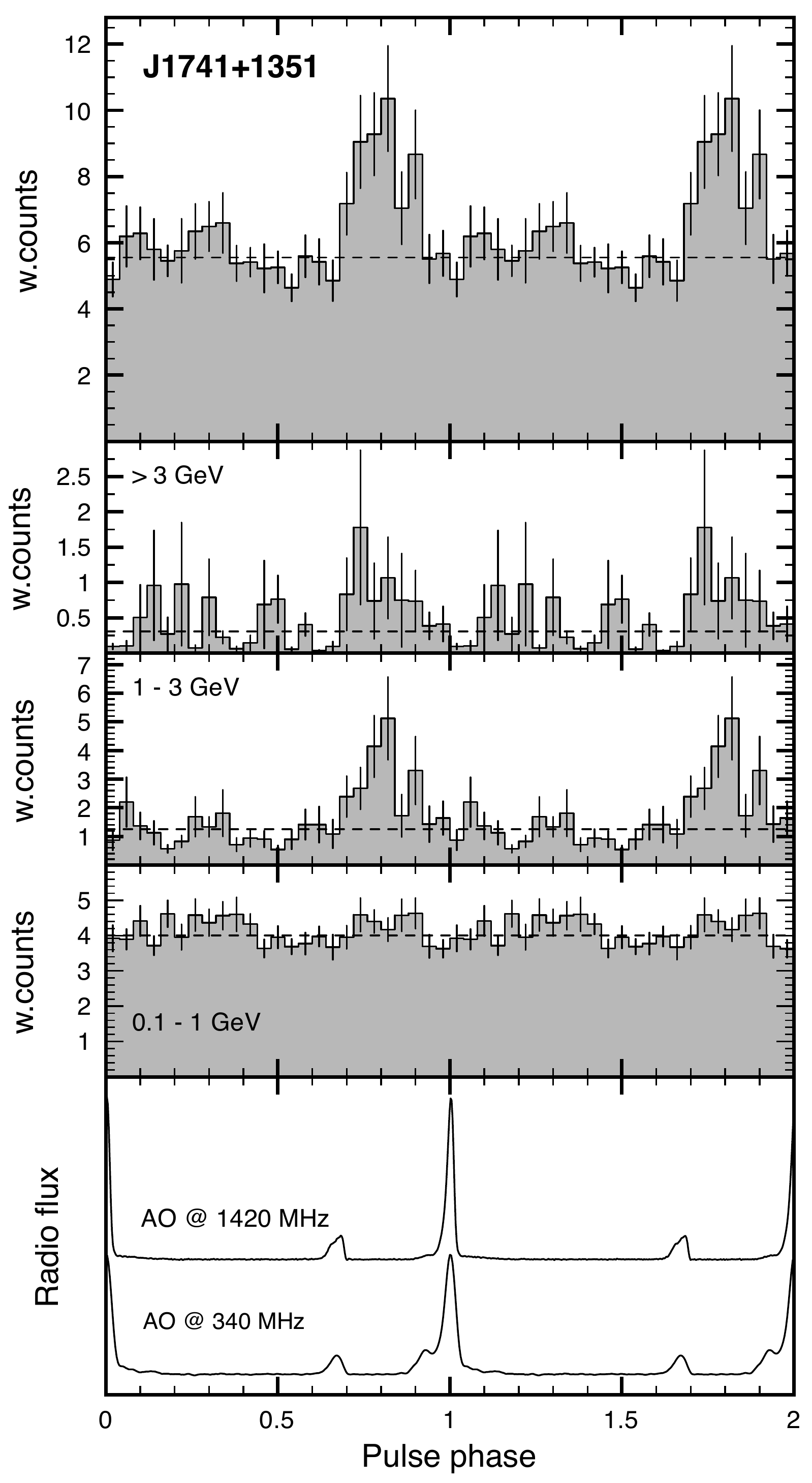}
\caption{Gamma-ray and radio pulse profile for PSR \sevenf\ (see Fig. \ref{0610lc} for details).}
\label{1741lc}
\end{center}
\end{figure}

The pulse profile of PSR \sevenf\ at $1.4$\,GHz exhibits one sharp main
pulse and a smaller 0.1 rotation wide additional component, leading the
main pulse by $\sim0.35$ rotations (see Fig. \ref{1741lc}).
There is also a small component on the base of the main pulse, which 
appears more prominent in the $0.3$\,GHz profile.
This component precedes the main pulse by $\sim0.05$ rotations.

\subsection{PSR J2051$-$0827}
This 4.5 ms pulsar is one of the so-called \emph{Black Widow} systems.
It is in a 2.4 hour orbit and exhibits radio eclipses 
for about 10\% of the orbital period, observable at low frequencies
\citep[$\le0.6$\,GHz,][]{sbl+96}.

\begin{figure}
\begin{center}
\includegraphics[width=75mm]{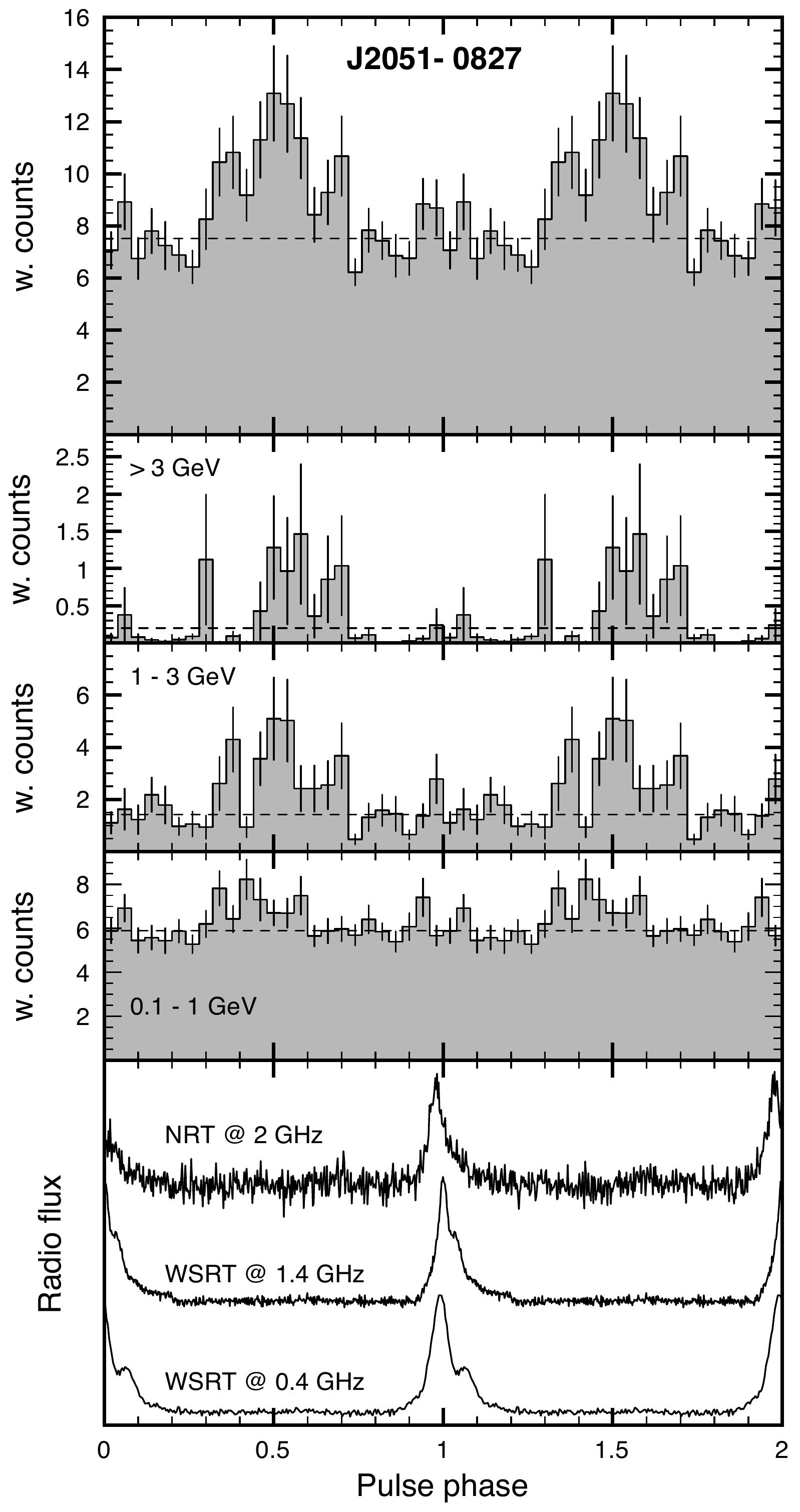}
\caption{Gamma-ray and radio pulse profile for PSR \twenty\ (see Fig. \ref{0610lc} for details).}
\label{2051lc}
\end{center}
\end{figure}

The gamma-ray profile of PSR \twenty\ exhibits one steep, $\sim0.4$ rotation wide, peak preceding the radio peak by half a period. 
The pulsed emission is similar in the three energy bands (Fig. \ref{2051lc}).
This profile is compatible with the light curve presented by \citet{wkh+12}.

\label{2051xrays}
XMM-{\it Newton} observed PSR \twenty\ in April 2009 with the
EPIC-MOS and EPIC-pn detectors, all operated in Full Window mode, for
44 and 37-ks effective exposures, respectively.  
Unfortunately, these observations suffered from numerous strong
particle flares that increased the background level by a factor of up
to 10-12 (compared to the normal one) during about 80\% of the
observations.  
This made the XMM-{\it Newton} data practically useless for a
purposeful analysis. 

Five observations of this object were conducted with the {\it Chandra}
ACIS-S instrument in Very Faint mode in March and July 2009 for $9$-ks
effective exposures \citep[see also][]{wkh+12}. 
The aim of these observations was to search for orbital variability of the pulsar X-ray flux and did not provide a time resolution suitable to perform a timing analysis at the pulsar spin period. 
We reduced the data using the CIAO software, version 4.3.
Despite the small number of source counts collected in each
observation ($8$--$12$ counts in a $1$\arcsec-radius aperture centred at the pulsar radio position), the object was clearly detected.
The combined dataset totalled 44 source counts in the $0.3$--$10$\,keV
range, with a negligible (less than 1\%) background contamination. 
The spatial distribution of these counts was found to be
consistent with the ACIS-S point-like source image.  
Of the extracted source counts, 90\% were detected at photon energies below 2 keV, indicating that the spectrum of PSR \twenty\ is soft
and likely of a thermal origin rather than of a non-thermal (magnetospheric) one. 
Indeed, fitting  a power-law model to the extracted spectrum resulted
in a large best-fitting photon index $\Gamma_{\rm X}\simeq 4$ and a
hydrogen column density $N_{\rm H}\simeq 2.3\times 10^{21}$\,cm$^{-2}$
 significantly greater than the value of $0.6\times 10^{21}$\,cm$^{-2}$
expected from the pulsar dispersion measure. 
Fixing $N_H=0.6\times 10^{21}$\,cm$^{-2}$ results in a power-law model with $\Gamma_{\rm X}\simeq 2.6$ and unabsorbed flux $F_{\rm X}\simeq 7.9\times 10^{-15}$\,erg\,cm$^{-2}$\,s$^{-1}$ in the $0.3$--$10$\,keV range, in agreement with the result reported by \citet{wkh+12}.  
The thermal blackbody model provided more reasonable parameters; 
$N_{\rm H}<1\times 10^{21}$\,cm$^{-2}$, a (redshifted) temperature
$T_{\rm pc}^\infty\simeq2.7\times 10^6$\,K and radius of the emitting
area (hot polar caps on the pulsar's surface) 
$R_{\rm pc}^\infty\simeq 0.04\,(d/1\,{\rm kpc})$\,km. 
The corresponding bolometric X-ray luminosity is 
$L_{\rm pl}^\infty\simeq 0.6\times 10^{30}\,(d/1\,{\rm 
  kpc})^2$\,erg\,s$^{-1}$.   
Using a non-magnetic hydrogen atmosphere model for thermal emission of MSPs
\citep{zps96,zav06,zav09}, we obtained
$T_{\rm pc}\simeq1.6\times 10^6$\,K, $R_{\rm pc}\simeq 0.18\,(d/1\,{\rm kpc})$\,km and 
$L_{\rm pc}\simeq 0.8\times 10^{30}\,(d/1\,{\rm kpc})^2$\,erg\,s$^{-1}$.
These are all unredshifted values, i.e. as measured at the neutron
star surface.  
 
Using the {\it photons} plugin for {\sc tempo2}, we searched for
orbital variability of the X-ray photons detected in the {\it Chandra} 
observations.
With the scant count statistics available, we found that the significance of a possible modulation does not exceed $2.2$\,$\sigma$. 
X-ray data of much better quality are required to elucidate the origin of the emission (whether it is thermal polar cap radiation or non-thermal flux from the pulsar magnetosphere and/or interaction of the pulsar wind with the companion).

The pulse profile of this MSP at $\sim1.4$\,GHz exhibits one main pulse
and one overlapping trailing component about 0.05 rotations apart. 
Additionally, there are at least 3 smaller trailing components
extending up to 0.2 rotations away, forming a shoulder.
The pulse profile at $0.4$\,GHz keeps the same principal structure but the two main peaks appear broader and slightly more separated.
There are no major differences with the profile at 2\,GHz (see
Fig. \ref{2051lc}). 
The emission at $1.4$\,GHz is mildly linearly polarised ($\sim$12\%) throughout the whole profile \citep{xkj+98}.

\section{Search for orbital modulation of the gamma-ray emission}
\label{orbsearch}
PSRs \osix\ and \twenty\ are in close binary orbits, where
interactions between the wind of the pulsar and the atmosphere of the
companion star are likely to occur.  Accordingly, we have searched for
gamma-ray flux modulations as these systems follow their orbital
cycles.  For completeness, we also searched for such modulations in
the other three binary systems, PSRs  \six, \seven\ and \sevenf.

The \fermi-LAT sensitivity to a source varies on multiple timescales,
most notably the spacecraft's orbital period ($\sim$95 minutes) and
period of precession ($\sim$53 days).  Beats between the binary
orbital frequency and harmonics of these timescales induce an
apparent modulation of photon rate as a function of orbital phase.
Searches for intrinsic modulation must therefore carefully correct for
the time-dependent sensitivity \citep[e.g.][]{ck10,gjv+12}.

Because of the relatively low count rates, we decided to adapt an
unbinned pulsation statistic, the H-test \citep{drs89}, for use with
uneven exposure.  To do this, we computed the exposure to the source
with $30$\,s resolution over broad energy bands (2 per decade).  We
used {\sc Tempo2} and the timing solutions of \S\ref{timing} to
compute the orbital phase for each $30$\,s interval and took the
resulting distribution of phases, $\mathcal{F}$, to represent the null
hypothesis of no intrinsic modulation.  By definition, the quantity
$\mathcal{F}(\phi)$, with $\phi$ the observed orbital phases, is a uniformly distributed random variable in the absence of intrinsic modulation, and thus suitable for use in the H-test.

Using this modified H-test, along with the photon weights (see
\S\ref{gammapart}), we searched for a signal in individual
energy bands (100--300, 300--1000, 1000--3000, 3000--10000, and
10000--30000\,MeV) as well as cumulatively ($>$100, $>$300,
$<$300\, MeV, etc.).  We detected no significant orbital modulation of
the gamma-ray signal from any of the sources.

\begin{figure}
\begin{center}
\includegraphics[width=75mm]{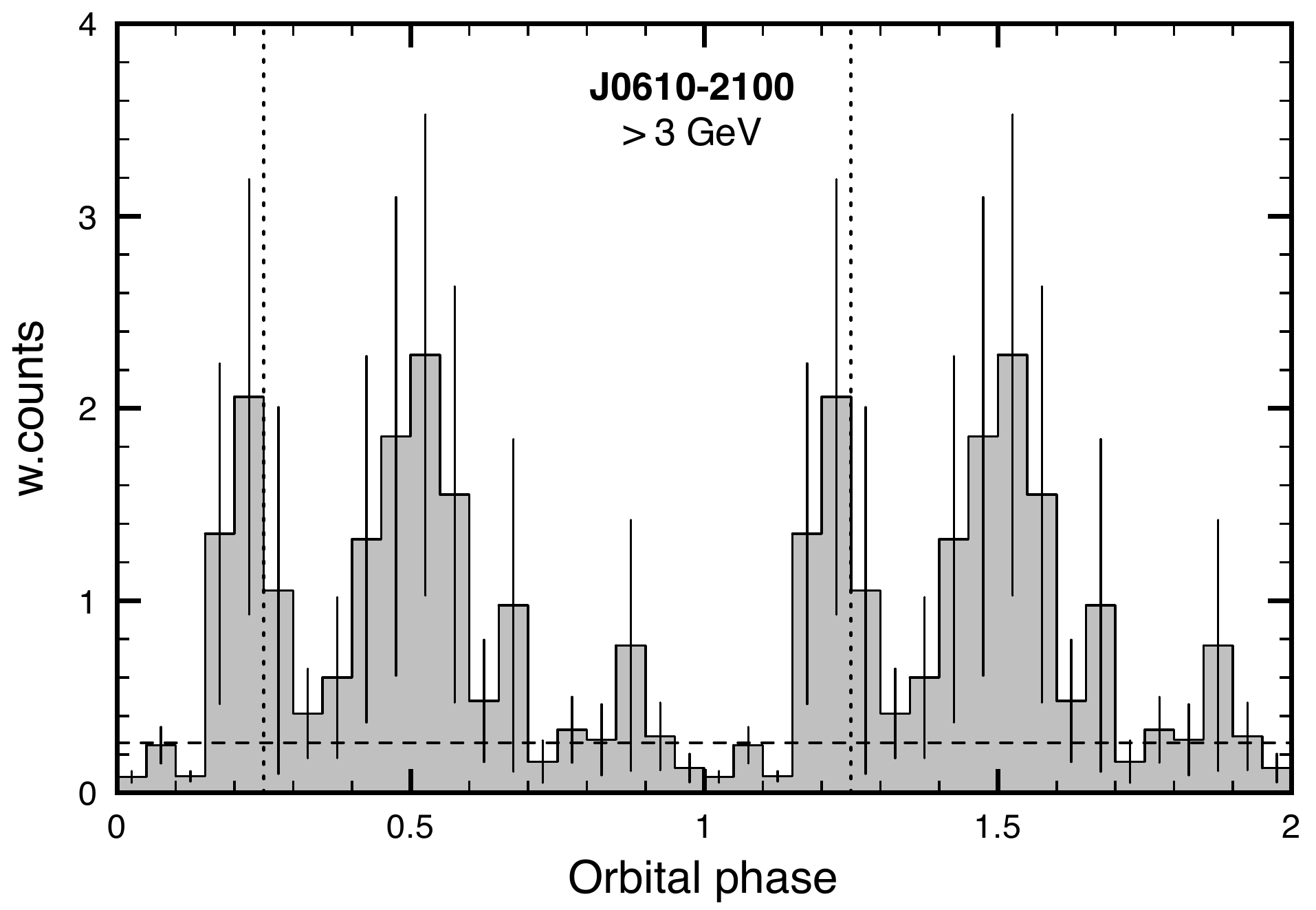}
\caption{Gamma-ray emission above $3$\,GeV folded on the orbital period for PSR \osix. 
	The light curve was made using photon weights and it is shown twice for clarity.
	Vertical dotted lines indicate the phase at which eclipses would be expected and the horizontal dashed line represents the photon background.	The H-test for this light curve is low, suggesting the modulation is not significant.}
\label{0610orb}
\end{center}
\end{figure}

This null result is physically expected for the binary systems with
long orbits, where no direct interaction between the two bodies is
feasible.  
In the case of PSR \twenty, this result is also
unsurprising as no evidence for interaction of the pulsar wind with
the companion star is seen in the X-ray spectrum (see section
\ref{2051xrays}).  Interestingly, the weighted orbital light curve at
energies above 3\,GeV for PSR \osix\ presents two peaks, suggesting
the presence of orbital modulation of the emission at these energies
(Fig. \ref{0610orb}).  However, there are very few events above
3\,GeV and the peaks observed could be the effect of low statistics;
indeed, the H-test for this set of events indicates only a
2\,$\sigma$ significance.  Furthermore, the light curve of the
background photons (obtained by assigning a weight $1-w_i$ to every
event in the same 5 deg ROI, see section 2.2.2.) for this source seems to follow the same shape in Fig. \ref{0610orb}.  
Although fairly flat, the exposure folded curve for
PSR \osix\ also presents local peaks at similar orbital phases ($\sim
0.2$ and $\sim 0.5$).

Finally, we performed a Monte Carlo study to determine the sensitivity
to a few particular types of modulation.  For each pulsar, we use the
observed weights and an assumed morphology for the orbital modulation
to generate 100 random realisations of the orbital phases.  We then
vary the strength of the modulation until 95 of the 100 simulations
exceed the statistical threshold, taken to be $H=8$, or a
$\sim$2$\sigma$ detection.

We find in general that only strong modulations are detectable.
Sinusoidal modulation is detectable for PSRs \osix\ and \seven\ if
$\sim$70\% of the total flux is modulated, while not even 100\%
sinusoidal modulation is reliably detectable for the remaining
sources.  On the other hand, such high levels of modulation, which
imply additional, non-magnetospheric sources of gamma-ray emission
from the system, are incompatible with the estimated background levels
of the rotational phase light curves (Figs. \ref{0610lc} to \ref{2051lc}).

For the case of a notch in otherwise-steady emission (an eclipse), we
find that the sensitivity depends strongly on the notch width.  Any
gamma-ray eclipse is expected to be narrow, and unfortunately the LAT
is largely insensitive to such small features.  Eclipses become
detectable for PSRs \osix\ and \seven\ when they span $\sim$20\% of
the orbit, while even greater values (30--50\%) are required for the
remaining sources.  Thus, while we can rule out extreme cases of
modulation (fully-modulated sinusoids, broad eclipses), our
sensitivity study shows the null results discussed above are not
highly constraining.

\section{Radio and gamma-ray properties of the gamma-ray MSP
  population} 
\label{morfanalys}
Our sample of \nsample\ MSPs is constituted by all MSPs known to emit in gamma rays at the time of writing (published or publicly announced). 
We have compiled published radio pulse profiles and measurements of
the radio spectral index ($\mathnormal{s}$), radio flux densities, average degree of radio linear polarisation through the main radio pulse, phase lag of the closest gamma-ray peak to the main radio peak ($\delta$), gamma-ray photon index ($\Gamma$) and flux for each of the \nsample\ MSPs in the sample.
Table \ref{ta:allmsps} gives some of these quantities and also lists references where gamma-ray and radio pulse profile plots can be found.

\begin{table*}
\begin{minipage}{130mm}
\caption{Radio and gamma-ray properties of \nsample\ MSPs detected by the LAT.
  \label{ta:allmsps}}
\begin{tabular}{lccccccl}
\hline
\multicolumn{1}{c}{Pulsar} & \multicolumn{1}{c}{$P$} & \multicolumn{1}{c}{\blc} &  \multicolumn{1}{c}{$\langle L/I\rangle$} & \multicolumn{1}{c}{$s$}  
& \multicolumn{1}{c}{$\delta$} & \multicolumn{1}{c}{Type} & \multicolumn{1}{l}{References}  \\
\multicolumn{1}{c}{} & \multicolumn{1}{c}{ms} & \multicolumn{1}{c}{$\times 10^{4}$\,G} & \multicolumn{1}{c}{\%}  & & \multicolumn{1}{c}{rotations} &   &   \\
\hline
J0030$+$0451	&	4.87	&	1.8	&	?      & $-2.2$(2)  &	0.160(1) &	N	&	1;	1;	21; 25  \\
J0034$-$0534	&	1.88	&	13.6	&	0(10)  & $-3$(1)    &	0.97(1)  &	A	&	2;	14;	22; 22	\\
J0101$-$6422	&	2.57	&	6.1	&	18(3)  & --         &	0.15(1)  &	N	&	3;	--;	21; 3	\\
J0218$+$4232	&	2.32	&	31.4	&	22(10) & $-2.8$(2)  &	0.71(2)  &	W	&	2;	15;	21; 25	\\
J0340$+$4130	&	3.30	&	4.1	&	--     & $-2.1$(7)  &	0.3(1)   &	N	&	--;	16;	30; 16	\\
J0437$-$4715	&	5.76	&	1.4	&	24(1)  & $-1.1$(5)  &	0.44(1)  &	N	&	4, 5;	14;	21; 25	\\
J0610$-$2100	&	3.86	&	1.2	&	--     & --         &	0.1(1)   &	N	&	--;	--;	21; 30	\\
J0613$-$0200	&	3.06	&	5.4	&	17(3)  & $-1.5$(5)  &	0.26     &	N	&	5, 6, 7; 14;	21; 25	\\
J0614$-$3329	&	3.15	&	7.0	&	--     & $-0.3$(6)  &	0.126(2) &	N	&	--;	17;	21; 26	\\
J0751$+$1807	&	3.48	&	3.6	&	29(2)  & $-0.9$(3)  &	0.40(1)  &	N	&	8;	18;	21; 25	\\
J1024$-$0719	&	5.16	&	$<0.3$	&	52(4) & $-1.5$(2)  &	0.47     &	N	&	5, 8, 6; 18;	21; 30	\\
J1125$-$5825	&	3.10	&	13.4	&	--     & --         &	0.6      &	N	&	--;	--;	23; 9	\\
J1231$-$1411	&	3.68	&	5.4	&	--     & --         &	0.24     &	N	&	--;	--;	21; 26	\\
J1446$-$4701	&	2.19	&	13.1	&	20(5)  & --         &	0.5      &	N	&	9;	--;	23; 9	\\
J1600$-$3053	&	3.60	&	3.5	&	30(4)  & $-0.6$(6)  &	0.16(2)  &	N	&	5, 6;	19;	21; 30	\\
J1614$-$2230	&	3.15	&	5.2	&	--     & --         &	0.19     &	N	&	--;	--;	21; 25	\\
J1713$+$0747	&	4.57	&	1.9	&	26(4)  & $-1.5$(1)  &	0.32(5)  &	N	&	5, 8, 6; 18;	21; 30	\\
J1741$+$1351	&	3.75	&	5.8	&	--     & --         &	0.76(2)  &	N	&	--;	--;	21; 30	\\
J1744$-$1134	&	4.08	&	2.3	&	90(4)  & $-1.8$(7)  &	0.82(1)  &   N$\dagger$	&	5, 6;	14;	21; 25	\\
B1820$-$30A	&	5.44	&	24.8	&	0(10)  & $-2.7$(9)  &	0.99(1)  &	A	&	2;	14;	21; 27	\\
J1902$-$5105	&	1.74	&	22.1	&	0(10)  & --         &	1.0      &	A	&	10;	--;	21; 10	\\
B1937$+$21	&	1.56	&	99.5	&	27(2)  & $-2.3$(2)  &	0.990(4) &	A	&	2, 5, 8, 11; 18; 24; 24	\\
B1957$+$20	&	1.61	&	25.2	&	0(3)   & $-3.5$(5)  &	0.99(2)  &	A	&	11, 12;	18;	24; 24	\\
J2017$+$0603	&	2.90	&	5.9	&	--     & --         &	0.21(1)  &	W	&	--;	--;	21; 28	\\
J2043$+$1711	&	2.38	&	8.0	&	--     & --         &	0.131(4) &	N	&	--;	--;	21; 29	\\
J2051$-$0827	&	4.51	&	2.4	&	12(1)  & $-1.6$(9)  &	0.51(4)  &	N	&	8, 6;	14;	21; 30	\\
J2124$-$3358	&	4.93	&	1.9	&	24(8)  & $-1.5$(8)  &	0.87(1)  &	W	&	5;	14;	21; 25	\\
J2214$+$3000	&	3.12	&	6.4	&	25(5)  & $-2.4$(5)  &	0.27(1)  &   N$\dagger$	&	13;	13;	21; 26	\\
J2241$-$5236	&	2.19	&	10.7	&	--     & $-0.8$(1.4)&	0.14(1)  &	N	&	--;	20;	21; 23	\\
J2302$+$4442	&	5.20	&	1.7	&	--     & --         &	0.45(1)  &	W	&	--;	--;	21; 28	\\
\hline
\end{tabular}
{\sc Note.---} Columns contain: pulsar name, magnetic field at the light cylinder (\blc), mean
degree of linear polarisation $\langle L/I\rangle$, radio spectral
index $s$, radio lag ($\delta$) and type of MSP, 
according to the classification in section \ref{morfanalys}. 
Uncertainties on the last quoted digit are indicated between parentheses.
For the average linear polarisation, the errors quoted correspond to a
rough and conservative estimate of the standard deviation from several
measurements made by different authors, preferably around $1.4$\,GHz. 
Type labels marked with the symbol $\dagger$ are for those MSPs with radio and gamma-ray pulses
overlapping in phase but not considered as aligned (section \ref{sct:atype}).
The 4 references in the last column are for $\langle L/I\rangle$, $s$,
$\delta$ and for a gamma-ray/radio pulse profile plot:
(1)		\citet{lzb+00};
(2)		\citet{stc99};
(3)		\citet{kcj+12};
(4)		\citet{nms+97};
(5)		\citet{ymv+11};
(6)		\citet{ovhb04};
(7)		\citet{mh04};
(8)		\citet{xkj+98};
(9)		\citet{kjb+12};
(10)         \citet{camilo12};
(11)		\citet{ts90};
(12)		\citet{fbb+90};
(13)			P. Demorest, S. Ramson (priv. communication)
(14)		\citet{tbms98};
(15)		\citet{nbf+95};
(16)			Bangale et al. ({\sl in preparation})
(17)			S. Ransom (priv. communication)
(18)		\citet{kxl+98};
(19)		\citet{dfg+12};
(20)			M. Keith (priv. communication)
(21)		\citet{Fermi2PC};
(22)		\citet{aaa+10b};
(23)		\citet{kjr+11};
(24)		\citet{gjv+12};
(25)          \citet{aaa+10c}
(26)          \citet{rrc+11}
(27)          \citet{faa+11}
(28)          \citet{cgj+11}
(29)          \citet{gfc+12}
(30)		\emph{This work};
\end{minipage}
\end{table*}

\subsection{Radio and gamma-ray pulse profile properties}
Considering properties like phase-alignment between radio and
gamma-ray peaks (above 0.1 GeV), radio duty cycle and number of pulse  components, we define three main groups and label them \aa, \bb\ and \cc\ (Table~\ref{ta:allmsps}). Although this analysis is based mainly on the inspection of average pulse 
profiles at $\sim1.4$\,GHz, we also inspected pulse profiles at lower
and higher radio frequencies. 
None of the MSPs studied exhibits appearance or disappearance of pulse components (though, see comments  on PSR B1957+20 below and on PSR J0218+4232 in section \ref{typB}).
We note that pulse-profile evolution with frequency is observed in some other MSPs.

\subsubsection{Type A(ligned)}
\label{sct:atype}
These are the MSPs which have their main gamma-ray peak aligned with the main radio pulse. 
There are \NA\ objects in this group (Table \ref{ta:allmsps}). 

All of the \aaa\ MSPs exhibit two main radio peaks, with each of them composed of one or more components. 
In two cases (PSRs B1937+21 and B1957+20) the second peak appears as an interpulse, i.e. half a rotation away from the main pulse. 
However, in the case of PSR B1937+21, low level emission detected between the peaks suggested that both peaks may not be due to emission produced at opposite magnetic poles but somewhere in the outer magnetosphere \citep{ymv+11}. 
Only \NA\ MSPs in the whole sample present secondary peaks about half a rotation away from the main pulse.

For the \NA\ \aaa\ MSPs, all radio peaks have a gamma-ray
counterpart.
The only exception might be PSR B1957+20, for which the second radio peak visible at $1.4$\,GHz, leading the main peak by $\sim 0.26$ rotation, appears not to have an obvious gamma-ray counterpart.
We note that this peak is not visible at lower radio frequencies (e.g. $0.3$\,GHz).

Four MSPs in this group exhibit very low or undetectable levels of linear polarisation in their radio emission (Table \ref{ta:allmsps}).
This seems to be a property exclusive to \aaa\ MSPs: all other gamma-ray MSPs in the sample for which polarisation data were available exhibit normal levels of linear polarisation.
The only \aaa\ MSP showing some degree of linear polarisation is PSR B1937+21.

\subsubsection{Type N(on-aligned)}
\label{typB}
These are MSPs having their main gamma-ray peak out of phase with the main radio peak and their radio emission dominated by a single peak. 
Two MSPs in this group present gamma-ray pulses almost phase-coincident with  their main radio pulses (PSRs J1744$-$1134 and J2214$+$3000).
However, they are not exactly aligned and their peaks lead the radio peak, which is unusual among gamma-ray pulsars; though it is also seen in the light curves of the MSPs PSR J1741+1351 and PSR J2124$-$3358. 
These objects may belong to a different class of MSPs, in terms of morphology \citep[cf.][]{Venter2009,joh11}.

There seems to be some gamma-ray emission, besides the main peak of emission, at the radio pulse phase in at least 3 objects (PSRs J0340+4130, J0610$-$2100 and J1024$-$0719). 
There are \NN\ \bbb\ MSPs in total, two of them exhibiting an
interpulse (PSRs J0030+0451 and J0101$-$6422).

\subsubsection{Type W(ide)}
This is a group of \NW\ MSPs that have their main gamma-ray peak out of
phase with the main radio peak and whose radio emission consist of multiple peaks of comparable amplitude, covering most of the rotational period.
PSR J0218+4232 presents a secondary radio pulse component coincident
with the gamma-ray peak of emission. 
This radio component is stronger at lower radio frequencies
($<0.6$\,GHz), becoming comparable in strength to the main radio peak
\citep{khv+02}. 
There also seems to be alignment between some secondary radio
components and the main gamma-ray peak for PSRs J2124$-$3358 and
J2302$+$4442. 
Nevertheless, they are not proper \aaa\ pulsars because not all of their radio peaks are aligned with gamma-ray peaks.
In addition, it must be noted that given the large number of radio components and their long duty cycles, these could be mere coincidences.  
Besides these practical reasons, there is a physical reason to
keep the W and N groups separated. 
The \ccc\ pulsars are possibly aligned rotators, i.e. pulsars with their magnetic axes almost coincident with their rotation axes, a situation that could affect their observed properties.

\begin{figure*}
\begin{center}
\includegraphics[width=100mm]{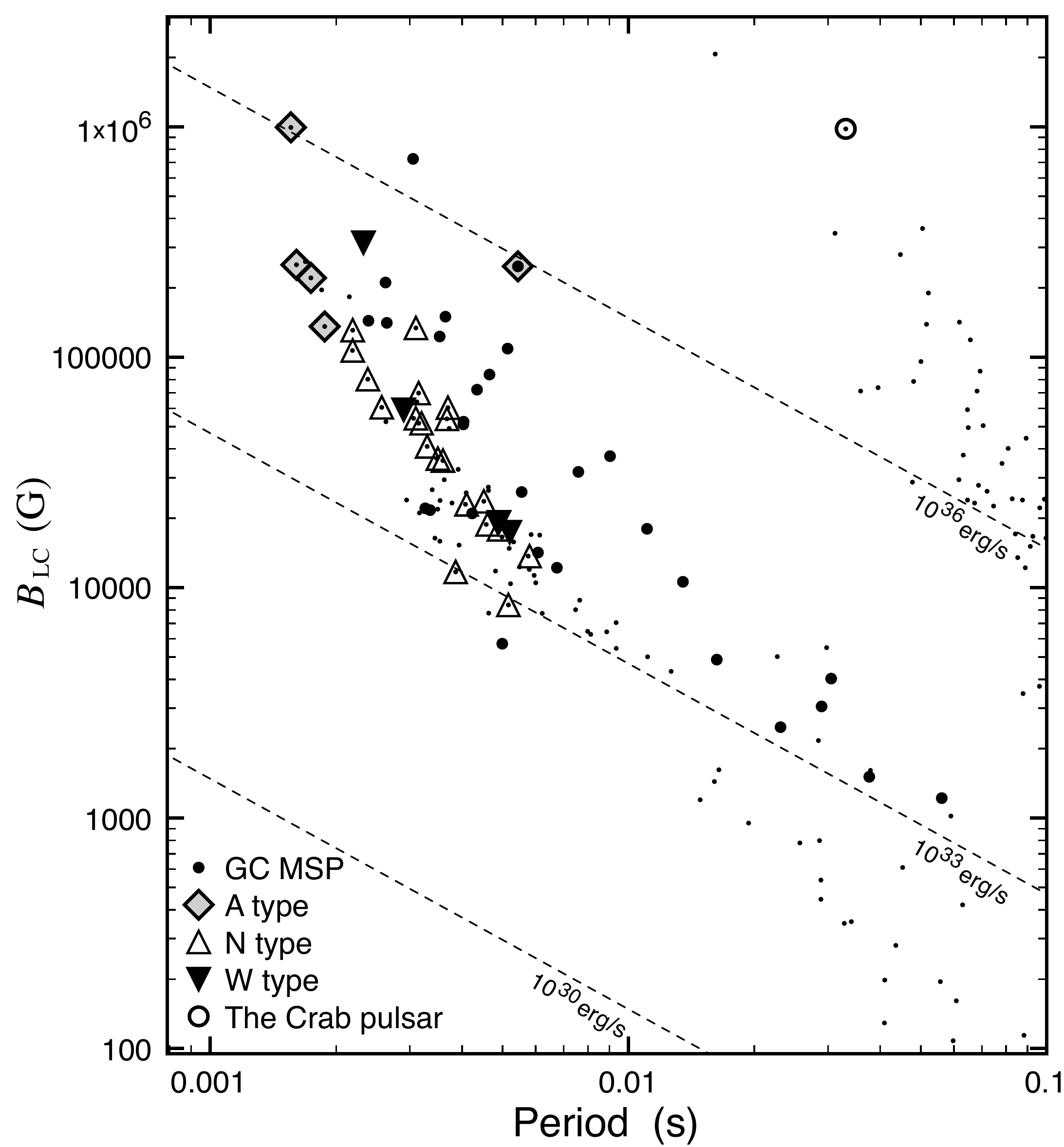}
\caption{Magnetic field at the light cylinder, corrected for the Shklovskii effect if possible, versus the rotational period for all known pulsars with $P\leq0.1$\,s. 
MSPs in globular clusters (GC) are shown using thick black dots to indicate that their \blc\ values are uncertain (see section \ref{sec:blc}). 
Lines of constant spin-down energy rate are dashed. 
}
\label{pbcyl}
\end{center}
\end{figure*}

\subsection{Magnetic field at the light cylinder}
\label{sec:blc}
The dipole magnetic field strength of a pulsar is normally calculated by assuming that all rotational energy losses are due to dipole radiation.
Its value at the light cylinder can be estimated by \citep[e.g.][]{lk05}
\begin{equation}
\label{blc}
B_\textsl{LC}=
   9.2\left(\frac{P}{1\,s}\right)^{-5/2}
   \left(\frac{\dot{P}}{10^{-15}}\right)^{1/2}~\textrm{G} \quad .
\end{equation}

At the light cylinder, the co-rotational velocity is equal to the
speed of light, $c$, and its radius is therefore given by $R_\textsl{LC}=cP/2\pi$, where $P$ is the rotation period.
This implies that gamma-ray MSPs have the smallest light cylinder radii, 
followed by rapidly spinning young pulsars like the Crab pulsar.
Because of the weak dependance of \blc\ on $\dot{P}$ and the different $\dot{P}$ ranges exhibited by MSPs and young pulsars, both populations share similar \blc\ values, the largest among the whole pulsar population ($>10^4$\,G).

\aaa\ MSPs tend to have higher \blc\ values 
than the rest of the population.
Fig.~\ref{pbcyl} shows \blc\ as a function of $P$ for all known pulsars
having $P\leq0.1$\,s. 
If proper motion measurements were available through the ATNF Pulsar
Catalogue, the \blc\ values were corrected for the Shklovskii
effect (Table \ref{ta:allmsps}).
Lines of constant spin-down energy rate $\dot{E}=4\pi^2 I \dot{P}/P^3$
(where $I$ is the moment of inertia of the star, assumed
to be $10^{45}$\,g\,cm$^2$) are plotted with dashed lines.
Most MSPs populate the lower branch crossing the centre of the plot
and the \nsample\ gamma-ray MSPs in our sample populate its higher
end, towards the shortest periods, and larger $\dot{E}$ and \blc\ values.
The MSPs from the three types described in the last section are plotted using different symbols.
\aaa\ MSPs are near the top left corner, with  the shorter periods and the higher \blc\ and $\dot{E}$ values \citep{joh11}.

PSR B1823$-$30A is one of the \aaa\ MSPs in the plot and, considering
its period, shows a relatively large \blc. 
This MSP is in a globular cluster (GC) and has a very large $\dot{P}$, hence its large \blc. \citet{faa+11} studied the gamma-ray emission
of this MSP and concluded that the observed $\dot{P}$ is intrinsic to
the pulsar and not due to acceleration in the cluster.
MSPs in GCs are marked in Fig.~\ref{pbcyl} and, due to possible contamination of their $\dot{P}$ values produced by movement in the gravitational potential of a GC, their \blc\ values are uncertain.

A 2-dimensional Kolmogorov-Smirnov (KS) test \citep{ff87,ptvf92}
between \aaa\ MSPs and \bb\ and \cc\ types combined
indicates that the probability that they belong to the same $P$--\blc\ distribution is less than 1\%.
However, the significance of the 2-dimensional KS test
remains valid while $N_e=N_1N_2/(N_1+N_2)\geq20$ \citep{ptvf92} and,
in this case, $N_e=4.2$. 
To check these results further, we also tried the standard
1-dimensional KS test, that can be applied for lower number of objects
\citep[valid for $N_e\geq4$,][]{ptvf92}.
Comparing the BLC-values of the five \aaa\
MSPs with  the 25 \bb\ and \cc\ types gives 0.02\%
probability that they belong to the
same distribution.
To quantify how likely it is to obtain such a value 
by selecting a small group at random, we randomly picked \NA\ MSPs from
the sample and calculated the KS test against the remaining \NNW\ objects.  
After repeating this 30,000 times we obtained that for more
than 99\% of the cases the KS test null hypothesis probability is
greater than 0.02\%, with 67\% of the cases having  a probability
greater than 40\%.
Therefore, we conclude that the distribution of \blc-values
of the \aaa\ MSPs is significantly different to that of the rest of
the gamma-ray MSPs.

We acknowledge, nonetheless, that the above results might be driven by the very short periods that most of the \aaa\ MSPs have, compared to those of the  \bb\ and \cc\ types together.
Indeed, while the KS test gives a probability of 41\% for the $\dot{P}$ values being from the same distribution, the same test gives only 0.4\% probability that the periods belong to the same distribution.
However, this is about 10 times larger than the probability obtained for \blc. 
If $\dot{E}$ values are considered instead, the probability is 1\%, which is again  50 times larger than what is obtained for \blc\ values.
Moreover, as will be discussed in section \ref{discussion}, the comparison with the Crab pulsar plus other considerations hint towards \blc\ being the relevant parameter, rather than $P$ or $\dot{E}$.

\subsection{Radio spectral indices}
\def \nspndx {19}  
\def \nmsps {52}   

\begin{figure}
\begin{center}
\includegraphics[width=80mm]{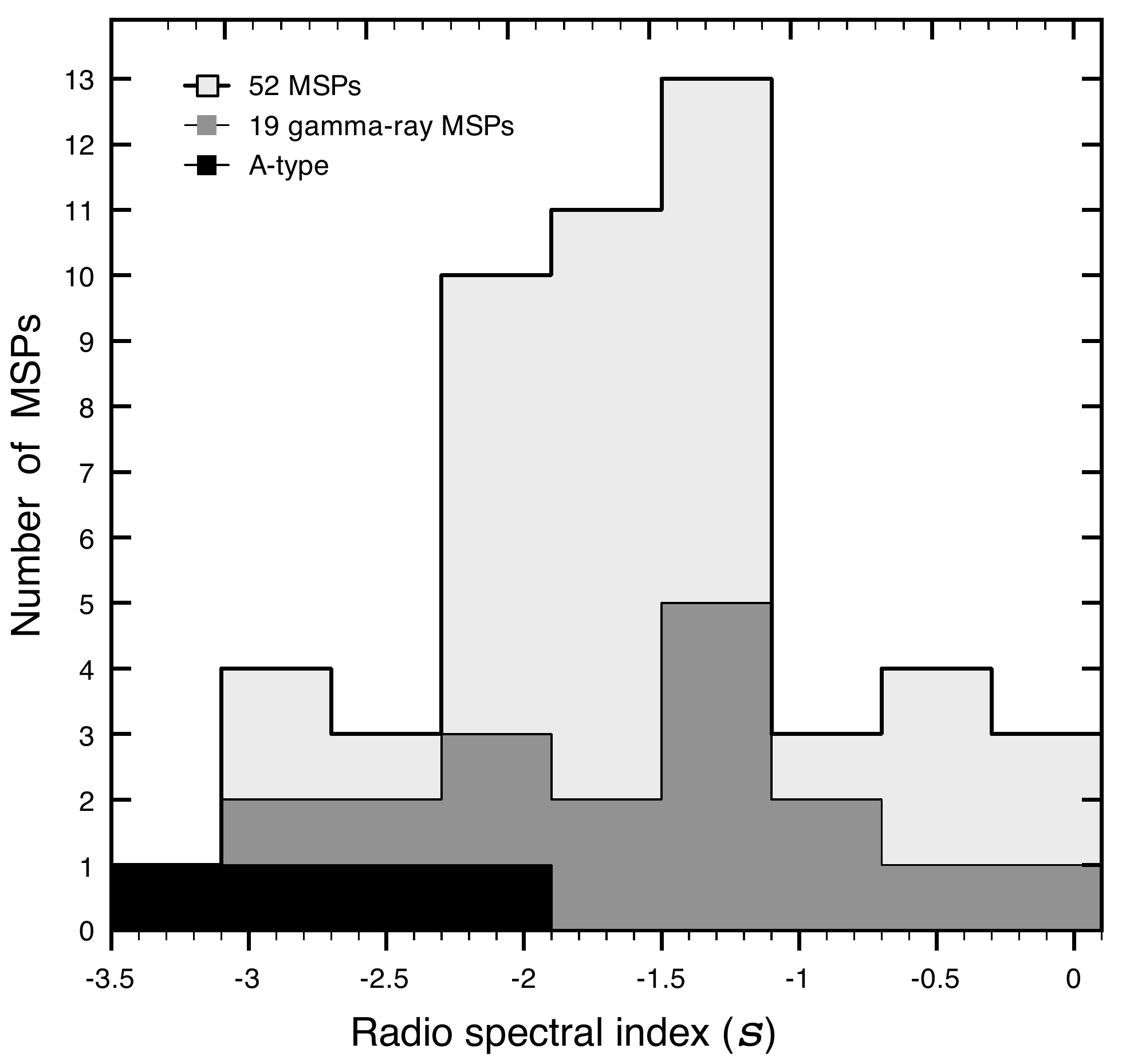}
\caption{Distribution of radio spectral indices
  for \nmsps\ MSPs highlighting the different values for gamma-ray MSPs
  and \aaa\ MSPs.}
\label{spIndxHisto}
\end{center}
\end{figure}

The radio spectral index $\mathnormal{s}$ corresponds to the slope of the spectrum in a logarithmic scale, in which the flux density is described by
$S\propto\nu^{\mathnormal{s}}$.
Out of the \nsample\ MSPs in the sample, we have spectral
information for \nspndx\ of them (Table \ref{ta:allmsps}).
These data show that \aaa\ MSPs tend to have lower spectral
indices than the rest of our sample.
Below we assess the significance of this observation.

The distribution of spectral indices for all the gamma-ray MSPs together with 33 additional MSPs for which spectral information was available is shown in Fig. \ref{spIndxHisto}, indicating the contribution made by the \aaa\ MSPs. 
Data were collected from the ATNF Pulsar
Catalogue, from the compilations in \citet{lylg95},
\citet{kxl+98} and \citet{tbms98}, and from flux densities at different
frequencies obtained via private communication with M. Keith and S. Ransom.
Note that the error bars of the $\mathnormal{s}$-values sometimes may not reflect the intrinsic difficulties associated with flux density measurements, which directly affect the spectral index estimate.
While the $\mathnormal{s}$ distribution of gamma-ray MSPs seems to follow the
general distribution, with a mean spectral index above $-2.0$
\citep[consistent with][]{kxl+98,tbms98}, the mean value for the
\aaa\ gamma-ray MSPs clearly falls below $-2.5$ (see
Fig. \ref{spIndxHisto}).

To test how significant the difference between the $\mathnormal{s}$
distributions of the different sets is, we again use the KS test.
A KS test over the $\mathnormal{s}$ values of four \aaa\ MSPs against the
remaining 15 gamma-ray MSPs (Table \ref{ta:allmsps}) indicates a
probability of 0.6\% that both sets belong to the same distribution.
If the test is applied for the \aaa\ MSPs against all other 
MSPs in Fig. \ref{spIndxHisto} the probability falls to 0.2\%.
Therefore, it appears that the \aaa\ MSPs have distinctively low
$\mathnormal{s}$-values. 
We note, however, that for the first KS test mentioned above,
$N_e=3.2$ (see section \ref{sec:blc}).
On the other hand, the test indicates a probability of 56.7\% that the
$\mathnormal{s}$ values of all gamma-ray MSPs belong to the overall
$\mathnormal{s}$ distribution, indicating that in general there is no
evidence for different radio spectral behaviour between gamma-ray
MSPs and the rest of the MSP population.

We also checked how likely it is to obtain an $\mathnormal{s}$
distribution such as the one exhibited by the \aaa\ MSPs by pure
chance. 
To do so, we randomly picked four values from the $\mathnormal{s}$-distribution of all gamma-ray MSPs and calculated the probability of them being from the same distribution as the remaining 15.  
After repeating this process many times, we find that more than 99\%
of the trials give a probability greater than 0.6\%, with more than
66\% of the trials giving a probability greater than 40\%.
Thus, the particular $\mathnormal{s}$-distribution of \aaa\ MSPs
appears to be truly different.

We find no correlation between $\mathnormal{s}$ and
the gamma-ray spectral index or photon index, $\Gamma$, for the MSPs in the sample.
Nonetheless, we note that values of $\Gamma$ are tightly clustered,
with a scatter smaller than unity \citep{aaa+10c}. 
As well, no obvious correlation was found between $\Gamma$ and any
other investigated parameter.

\subsection{Fluxes}
There are radio flux densities at $1.4$\,GHz available for 23 of the \nsample\ MSPs in our sample and gamma-ray fluxes for all of them.
These last values are preliminary results of the ongoing effort to produce the second \fermi\ LAT Catalog of gamma-ray pulsars by the \fermi\ collaboration \citep{Fermi2PC}.
For many MSPs in the sample, flux values can also be found in the references given for the gamma-ray light curves in Table \ref{ta:allmsps}.
We find no obvious correlation between radio flux densities and gamma-ray fluxes \citep[][]{aaa+12b}.

Fluxes are expected to be at least slightly correlated because of
their mutual dependance on distance.
The fact that this correlation is not observed in the sample suggests
that radio and/or gamma-ray fluxes are highly dependent upon
geometrical factors and probably other intrinsic properties dependant
on $P$ and $\dot{P}$. 
No correlation between these fluxes, or their ratio, and any
other quantity was found.

\subsection{Linear polarisation}
\label{sct:pol}
\begin{figure}
\begin{center}
\includegraphics[width=80mm]{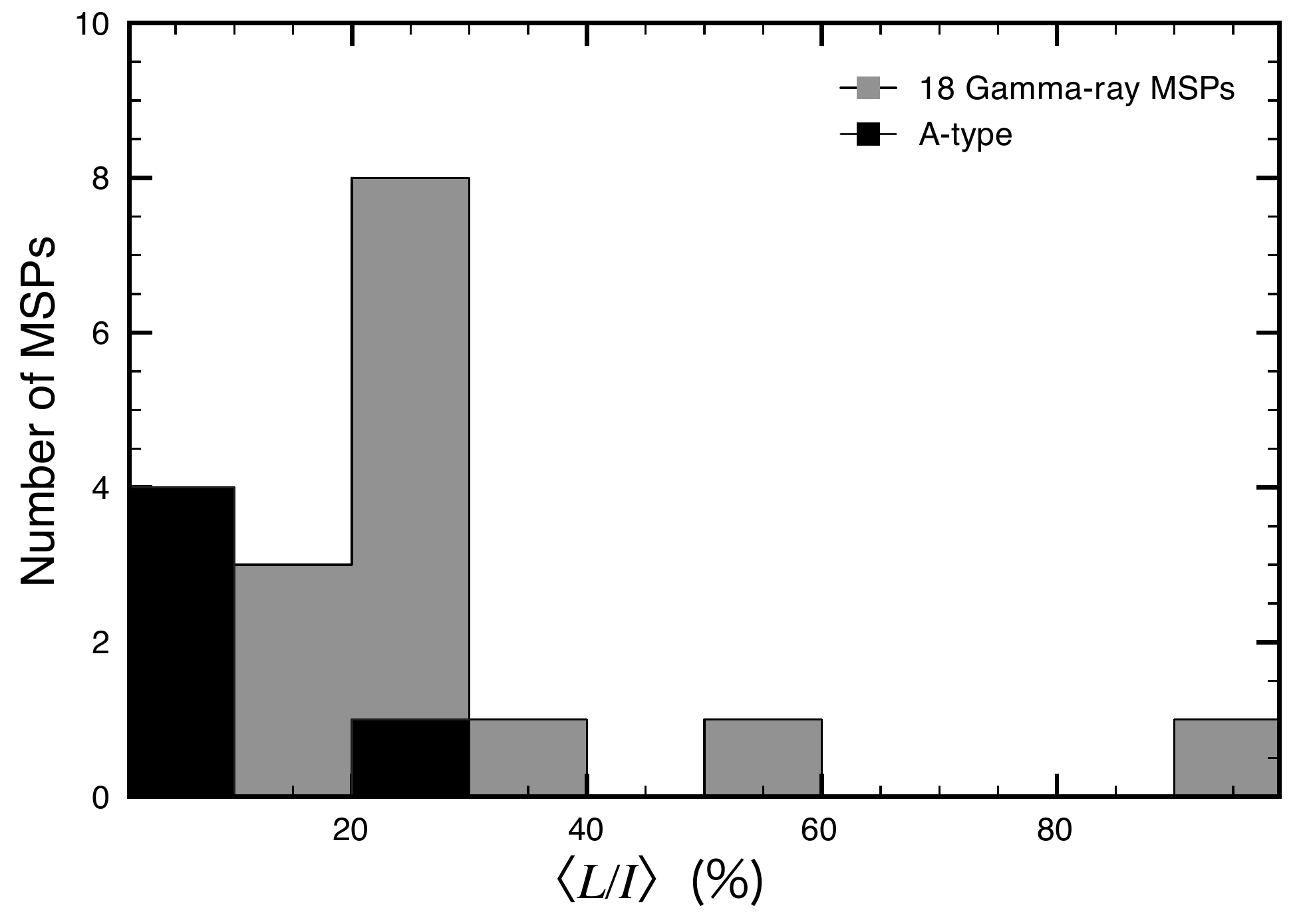}
\caption{Distribution of the average levels of radio linear polarisation for 18 MSPs in the sample.}
\label{polsHisto}
\end{center}
\end{figure}

\def \npol {17}
We found published average radio linear polarisation $\langle
L/I\rangle$ data for \npol\ MSPs in our sample.
No correlation is found between this parameter and any other studied quantity. 
However, we note that the four MSPs with very low (or zero) $\langle L/I\rangle$ are all \aaa\ (PSRs J0034$-$0534, B1820$-$30A, J1902$-$5105 and B1957+20, see Fig. \ref{polsHisto}).

\section{Discussion}
\label{discussion}

\subsection{Gamma-ray MSPs at a glance}
By studying the radio and gamma-ray pulse profiles of \nsample\ MSPs
we have classified gamma-ray MSPs into 3 types, labelled \aa, \bb\ and \cc.
The first group is composed by MSPs exhibiting pulsed gamma-ray emission
phase-aligned with their radio emission. 
The \bb\ and \cc\ type MSPs exhibit misaligned emission, with the \bbb
s presenting radio emission dominated by a single pulse and the \ccc\
presenting multiple and wide radio pulses of comparable amplitude.
While the \bb\ and \cc\ types appear to constitute the norm among gamma-ray
MSPs, \aaa\ MSPs appear less frequently and present various notable
differences from the rest of the population.
We note that these properties are not shared by PSRs J1744$-$1134 and J2214+3000, the two MSPs which could have possibly been regarded as \aaa\ (see sections \ref{sct:atype} and \ref{typB}).

\ccc\ MSPs are more likely to be aligned rotators, with their rotation
axes almost coincident with their magnetic axes. 
This situation, however, does not appear to drive obvious trends
either in their gamma-ray profiles or any of the other studied parameters.

The difference in rotational phase between radio and gamma-ray pulses
has commonly been interpreted as a difference in the location, within
the pulsar's magnetosphere, where the emission is generated.
Standard models normally place the source of radio emission at lower
altitudes compared to the source of gamma rays, in accordance with the
radio/gamma-ray misalignment observed for most gamma-ray pulsars and
MSPs.  
However, in light of the newly detected MSPs exhibiting phase-aligned 
emission, other possibilities ought to be considered.
In the two most likely scenarios, both radio and gamma-ray emission
are generated either relatively near the surface of the neutron star
(low-altitude Slot Gap (laSG) models) or over an extended region at
higher altitudes (altitude-limited Open Gap (alOG) and altitude limited
Two Pole Caustic (alTPC) models \citep[cf.][]{vjh12}). 
Although the alTPC model seems to offer better fits for some \aaa\ MSPs, further studies are necessary to better understand the emission mechanisms operating in these pulsars.

Nonetheless, models involving a caustic origin for the radio emission
(at high altitudes) predict low levels of linear polarisation and
rapid swings of the polarisation position angle \citep{dhr04}.
In such models, rapid position angle swings with phase and
depolarisation of the total emission would be an effect of mixing
signals from different altitudes \citep{vjh11}.
This is indeed observed in some, but not all, of the \aaa\ MSPs
(section \ref{sct:pol}; Table \ref{ta:allmsps}). 

It has been noted that for high $\dot{E}$ non-MSP pulsars the degree of
linear polarisation may be correlated with $\dot{E}$ \citep{wj08}.
Moreover, based on their wide beams, it was argued that the
radio emission of some of these pulsars is produced at a large range
of heights.
Their polarisation levels, however, appear high, contrary to what is
argued above to explain the lack of polarised emission in some \aaa\
MSPs. 
Also, \aaa\ MSPs have the largest $\dot{E}$ values among MSPs,
defying the correlation found by \citet{wj08}.
Nevertheless, normal pulsars have larger light cylinder radii than MSPs and it might be that depolarisation is less efficient in their larger magnetospheres.

In the smaller magnetospheres of MSPs the production of
radio emission could naturally extend up to larger fractions of
$R_\textsl{LC}$, producing the observed wide beams \citep{ymv+11} and the phase-aligned profiles of \aaa\ MSPs \citep{har05a}.
In fact, \aaa\ MSPs have shorter rotation periods than the
other MSPs (Fig. \ref{pbcyl}).
Based on a study of the beaming fractions of the radio and gamma-ray
emission, \citet{rmh10} concluded that the radio emission of
high-$\dot{E}$ pulsars (including MSPs) must originate higher up in the
magnetosphere, near where the gamma-ray emission is produced.
This is consistent with the high-$\dot{E}$ values of \aaa\ MSPs.
The Crab pulsar is the only non-MSP known to exhibit
nearly phase-aligned radio and gamma-ray emission \citep{khwf03}. 
We note that $\dot{E}$ for the Crab pulsar is almost 3
orders of magnitude larger than the highest values among MSPs and that
many young gamma-ray pulsars have $\dot{E}$ values similar to those of the \aaa\ MSPs, but show misaligned emission.
Also, while the Crab pulsar has the second smallest $R_\textsl{LC}$
value among normal pulsars, it is more than one order of magnitude larger than those of the \aaa\ MSPs.
If $R_\textsl{LC}$ was the main factor determining the
alignment between radio and gamma-ray emission, we would expect to see
this alignment for most MSPs, which is not observed.
On the other hand, if $\dot{E}$ was the main factor, we would expect
to see more young gamma-ray pulsars exhibiting aligned radio/gamma-ray
emission, which is not the case. 
Considering the KS-tests described in section \ref{sec:blc} and the position of the Crab pulsar in Fig. \ref{pbcyl}, \blc\ appears to be a natural common property among \aaa\ MSPs and the Crab pulsar.
We note that the polarisation levels of the Crab pulsar's
radio emission appear to be well above zero \citep{gl98}.

The Crab pulsar and \aaa\ MSPs all have similar radio spectra.
The radio spectral index of the Crab pulsar is $-3.1$ \citep{lylg95},
substantially steeper than the average of $-1.9$ for MSPs and of $-1.8$
for normal pulsars \citep{tbms98,mkkw00a}. 
Although viewing angles and other geometric factors can bias our
measurements, the steep radio spectra of the \aaa\ MSPs and the Crab
pulsar, together with their high \blc\ values, are likely to be
related to magnetospheric similitudes and common processes on the 
generation of their emission.

The emission of Giant radio Pulses (GPs) is another common feature among \aaa\ MSPs and the Crab pulsar.
GPs are sporadic, short and intense bursts of radio
emission, following power-law energy statistics.
They were originally detected in the emission of the Crab pulsar and
later in PSR B1937+21 \citep{hcr70,cstt96}.
Today there are 8 pulsars and 5 MSPs known to emit these type of pulses \citep{kbmo05,kni06}.
The GP properties exhibited by the Crab pulsar differ from most other pulsars but are very similar to those exhibited by the 5 MSPs found to emit GPs \citep{kni06}.
Three of these MSPs are \aaa: PSRs B1937+21, B1957+20 \citep{kbm+06}  and B1820-30A \citep{kbmo05} and the other two are PSRs J0218+4232 \citep{jkl+04}, a \ccc\ MSP, and B1821$-$24 \citep{rj01}, for which no gamma-ray pulses have been detected with a confidence above 5\,$\sigma$ \citep{ppp+09}.
It has been proposed that the relatively large \blc\ values exhibited
by these 5 MSPs and the Crab pulsar could be the main physical factor determining the generation and the main properties of their GPs \citep[][but see discussion in \citet{kbm+06}]{cstt96}.
Future observations and careful analyses should discern whether the emission of GPs is somewhat connected to the alignment between radio and gamma-ray emission or is mere coincidence.
No GPs were detected for the \aaa\ PSR J0034$-$0534
\citep{kbmo05}. 

The wide pulse profiles that MSPs generally have, compared to normal pulsars \citep{ymv+11}, could be understood as evidence for outer-magnetosphere caustic radio emission for all types of MSPs. 
One could argue that the outer magnetosphere offers enough room for different emission locations for radio and gamma rays.
However, we have shown that A-type MSPs exhibit different emission properties.
Could the availability of larger magnetic fields at the light cylinder
generate conditions, in relatively smaller magnetospheres, that favour
the generation of radio emission at higher altitudes (at least as a fraction of $R_\textsl{LC}$) and co-located with the production of gamma rays?
Would this somewhat different emission mechanism naturally produce a
steeper spectra and favour the production of giant pulses?
Any model describing phase-aligned radio and gamma-ray emission should take these properties into consideration.

\section{Summary}
\label{summary}

We have presented the detection by the \fermi\ LAT of gamma-ray
pulsations from \detected\ MSPs, five of them detected for the first time and a
sixth one, PSR \twenty, confirmed at the 5\,$\sigma$ level.
The \detected\ MSPs present properties which are common among the gamma-ray MSP 
population. 
All of these pulsars are significantly detected in the radio domain, but due to the low conversion of spin-down energy into X-rays, only two are detected significantly in X-rays.

By studying the morphology and phase relationship of radio and gamma-ray pulse profiles of a sample of \nsample\ MSPs, we grouped gamma-ray MSPs
into three types.
The most distinctive type of gamma-ray MSPs are those exhibiting
phase-aligned radio/gamma-ray emission.
We find some clear trends in their emission properties, which differ
significantly from the rest of the MSP population.
We showed that the MSPs in this group have a radio spectra steeper than
the rest of the MSP population and that they also have among the highest
inferred magnetic field strengths at the light cylinder.
Additionally, some MSPs in this group have distinctively low degrees of radio linear polarisation and some of these type are amongst the handful of MSPs known to emit giant radio pulses.
Many of these properties are also observed from the Crab pulsar, the only
normal radio pulsar known to emit gamma-ray emission phase-aligned
with its radio emission.

The use of combined information obtained through the study of phase aligned gamma-ray, X-ray and radio emission, along with their intrinsic properties, offers a wide perspective that certainly helps to
improve our understanding of the emission mechanism of pulsars.

\section*{Acknowledgments}
The \textit{Fermi} LAT Collaboration acknowledges generous ongoing
support from a number of agencies and institutes that have supported
both the development and the operation of the LAT as well as
scientific data analysis. 
These include the National Aeronautics and Space Administration and
the Department of Energy in the United States, the Commissariat \`a
l'Energie Atomique and the Centre National de la Recherche
Scientifique / Institut National de Physique Nucl\'eaire et de
Physique des Particules in France, the Agenzia Spaziale Italiana and
the Istituto Nazionale di Fisica Nucleare in Italy, the Ministry of
Education, Culture, Sports, Science and Technology (MEXT), High Energy
Accelerator Research Organization (KEK) and Japan Aerospace
Exploration Agency (JAXA) in Japan, and the K.~A.~Wallenberg
Foundation, the Swedish Research Council and the Swedish National
Space Board in Sweden. 
Additional support for science analysis during the operations phase is
gratefully acknowledged from the Istituto Nazionale di Astrofisica in
Italy and the Centre National d'\'Etudes Spatiales in France. 

The Nan\c cay Radio Observatory is operated by the Paris Observatory,
associated with the French Centre National de la Recherche
Scientifique (CNRS). The Lovell Telescope is owned and operated by the
University of Manchester as part of the Jodrell Bank Centre for
Astrophysics with support from the Science and Technology Facilities
Council of the United Kingdom. The Westerbork Synthesis Radio
Telescope is operated by Netherlands Foundation for Radio Astronomy,
ASTRON. 
The Parkes radio telescope is part of the Australia Telescope which is
funded by the Commonwealth Government for operation as a National
Facility managed by CSIRO. We thank our colleagues for their
assistance with the radio timing observations. The Arecibo Observatory is operated by SRI International under a cooperative agreement with the National Science Foundation (AST-1100968), and in alliance with Ana G. M\'endez-Universidad Metropolitana, and the Universities Space Research Association.

\bibliographystyle{mn2e-mod}
\bibliography{journals,modrefs,psrrefs,crossrefs,nuevas,biblio}

\label{lastpage}

\end{document}